\documentclass[journal]{IEEEtran}
\usepackage{amssymb,amsmath}
\usepackage[dvips]{graphicx}
\usepackage[usenames]{color}
\usepackage{amsfonts}
\usepackage{bm}
\usepackage{subfigure}
\usepackage{booktabs,multirow}
\usepackage{color}
\usepackage{setspace}

\IEEEoverridecommandlockouts

\begin{document}
%\markboth{IEEE TRANSACTION ON WIRELESS COMMUNICATIONS, Vol. XX,
%No. XX, Month 2015} {Ge etc.: Multi-user Massive MIMO Communication Systems Based on Irregular Antenna Arrays \ldots}
\title{\mbox{}\vspace{0.40cm}\\
\textsc{Multi-user Massive MIMO Communication Systems Based on Irregular Antenna Arrays} \vspace{0.2cm}}

\author{\normalsize
Xiaohu Ge,~\IEEEmembership{Senior Member,~IEEE}, Ran Zi,~\IEEEmembership{Student Member,~IEEE},
Haichao Wang, Jing Zhang,~\IEEEmembership{Member,~IEEE},

Minho Jo,~\IEEEmembership{Member,~IEEE}\\
%\vspace{0.70cm} \small{
%$^1$School of Electronic Information and Communications\\
%Huazhong University of Science and Technology, Wuhan 430074, Hubei, P. R.
%China.\\}
%Email: \{xhge, ranzi\_pds, m201371755, zhangjing\}@mail.hust.edu.cn\\
%\vspace{0.1cm}
%$^2$Department of Computer and Information Science, \\
%Korea University, S. Korea.\\
%Email: minhojo@korea.ac.kr\\
%\vspace{0.2cm}

\thanks{Xiaohu~Ge, Ran~Zi, Haichao~Wang and Jing~Zhang are with the School of Electronic Information and Communications, Huazhong University of Science and Technology, Wuhan 430074, Hubei, P. R. China. Emails: \{xhge, ranzi\_pds, wanghaichao, zhangjing\}@hust.edu.cn.}

\thanks{Minho~Jo is with the Department of Computer and Information Science, Korea University, Seoul 136-701, South Korea. Email: minhojo@korea.ac.kr.}

\thanks{Correspondence author:  Dr. Jing Zhang and Minho Jo, Tel: +86 (0)27 87557942, Fax: +86 (0)27 87557943, Email: zhangjing@mail.hust.edu.cn, minhojo@korea.ac.kr.}
\thanks{The authors would like to acknowledge the support from the National Natural Science Foundation of China (NSFC) under the grants 61210002 and 61271224, the Hubei Provincial Science and Technology Department under the grant 2013BHE005, the Fundamental Research Funds for the Central Universities under the grant 2015XJGH011 and 2014TS100. This research is partially supported by EU FP7-PEOPLE-IRSES, project acronym S2EuNet (grant no. 247083), project acronym WiNDOW (grant no. 318992) and project acronym CROWN (grant no. 610524). This research is supported by the National international Scientific and Technological Cooperation Base of Green Communications and Networks (No. 2015B01008) and Hubei International Scientific and Technological Cooperation Base of Green Broadband Wireless Communications.}
}

%\vspace{1cm}
%\underline{{$^{^\dagger}$}Corresponding Author's Address:}\\
%$\mbox{Cheng-Xiang Wang}$\\
%Joint Research Institute for Signal and Image Processing\\
%School of Engineering \& Physical Sciences\\
%Heriot-{\binom{C}{n}}Watt University\\
%Edinburgh EH14 4AS, UK\\
%Tel: +44 131 4513321\\
%Fax: +44 131 4514155\\
%Email: {\tt cheng-xiang.wang@hw.ac.uk}}
%\date{\today}
%\renewcommand{\baselinestretch}{1.2}
%\thispagestyle{empty} \maketitle \thispagestyle{empty}
%\newpage
%\setcounter{page}{1}
\maketitle

\begin{abstract}

In practical mobile communication engineering applications, surfaces of antenna array deployment regions are usually uneven. Therefore, massive multi-input-multi-output (MIMO) communication systems usually transmit wireless signals by irregular antenna arrays. To evaluate the performance of irregular antenna arrays, the matrix correlation coefficient and ergodic received gain are defined for massive MIMO communication systems with mutual coupling effects. Furthermore, the lower bound of the ergodic achievable rate, symbol error rate (SER) and average outage probability are firstly derived for multi-user massive MIMO communication systems using irregular antenna arrays. Asymptotic results are also derived when the number of antennas approaches infinity. Numerical results indicate that there exists a maximum achievable rate when the number of antennas keeps increasing in massive MIMO communication systems using irregular antenna arrays. Moreover, the irregular antenna array outperforms the regular antenna array in the achievable rate of massive MIMO communication systems when the number of antennas is larger than or equal to a given threshold.

\end{abstract}

%\IEEEpeerreviewmaketitle \vspace{-1cm}

\section{Introduction}
\label{sec1}

 To meet the challenge of 1000 times wireless traffic increasing in 2020 as compared to the wireless traffic level in 2010, the massive multi-input-multi-output (MIMO) technology is presented as one of the key technologies for the fifth generation (5G) wireless communication systems \cite{Ge16,Ge14,Chen15}. Existing studies validated that massive MIMO systems can improve the spectrum efficiency to 10-20 bit/s/Hz level and save 10-20 times energy in wireless communication systems \cite{Pitarokoilis12}. However, considering a limited available physical space for deployment of large number of antenna elements in base stations (BSs), the mutual coupling effect among antenna elements is inevitable for massive MIMO wireless communication systems \cite{Xu10}, \cite{Svantesson01}. Moreover, with hundreds of antennas deployed, new issues of the antenna array deployment and architecture may appear \cite{Boccardi14}. For example, conformal antenna arrays on the surface of buildings may no longer have uniform antenna spacings because of uneven surfaces of the deployment region. In this case, the antenna array becomes irregular and then the impact of mutual coupling on the massive MIMO system is different from the case with a regular antenna array. Therefore, it is an interesting and practically valuable topic to investigate multi-user massive MIMO communication systems using irregular antenna arrays.

%\subsection{Related Work}

When antennas are closely deployed in an antenna array, the interaction between two or more antennas, i.e., the mutual coupling effect, is inevitable and affects coefficients of the antenna array \cite{Balanis12}. The mutual coupling effect has been widely studied in antenna propagation and signal processing topics \cite{Andersen76,Gupta83}. Based on the theoretical analysis and experimental measurement, the performance of antenna array was compared with or without the mutual coupling effect in \cite{Andersen76}. It was shown that the mutual coupling significantly affects the performance of adaptive antenna arrays with either large or small inter-element spacing because the steering vector of the antenna array has to be modified both in phases and amplitudes \cite{Gupta83}.
With the MIMO technology emerging in wireless communication systems, the impact of mutual coupling on MIMO systems has been studied \cite{Kildal04,Janaswamy02,Clerckx07,Wallace04}. In a reverberation chamber, measurements and simulation results showed that the mutual coupling increases the spatial correlation level and undermines the MIMO channel estimation accuracy as well as the channel capacity \cite{Kildal04}. Considering a fixed-length linear array that consists of half-wave dipoles, simulation results revealed that the mutual coupling leads to a substantially lower capacity and reduces degrees of freedom in wireless channels \cite{Janaswamy02}. Moreover, analytical results in \cite{Clerckx07} showed that in a $2 \times 2$ MIMO system, the mutual coupling is detrimental to the subscriber unit (SU) correlation and simultaneously beneficial to the channel energy only in the presence of directional scattering conditions and for SU arrays oriented orthogonally to the main direction of arrival with spacings between 0.4 and 0.9 wavelengths.
Based on the scattering parameter matrix and power constraint, a closed-form capacity expression of the MIMO system with mutual coupling was derived \cite{Wallace04}.
%A matching network approach was proposed to compensate the disadvantages caused by mutual coupling on compact MIMO systems \cite{Lau06}. For broadband systems with uniform circular arrays, a unified communication-theoretic framework was presented to evaluate diversity limits and an optimal broadband matching network was derived to improve the performance of MIMO systems \cite{Taluja13}.
However, all of the above studies are based on conventional MIMO systems, i.e., antennas at the transmitter and receiver are less than or equal to $8 \times 8$. For future massive MIMO scenarios with hundreds of antennas placed at the BS, the mutual coupling effect needs to be further investigated.

To further improve the transmission rate in 5G wireless communication systems, the massive MIMO technology is envisaged to satisfy 1000 times wireless traffic increase in the future decade \cite{Marzetta10,Ngo13,Chen15_2}. Marzetta revealed that all effects of uncorrelated noise and fast fading will disappear when the number of antennas grows without limit in wireless communication systems \cite{Marzetta10}. Moreover, massive MIMO systems could improve the spectrum efficiency by one or two orders of magnitude and the energy efficiency by three orders of magnitude for wireless communication systems \cite{Ngo13,Mohammed14}. New precoding and estimation schemes have also been investigated for massive MIMO systems \cite{Chen14,Duly14}. Motivated by these results, the impact of mutual coupling on massive MIMO systems was explored in recently literatures \cite{Shen10,Artiga12,Masouros13}. For massive MIMO systems where dipole antennas are placed in a fixed length linear array, analytical results indicated that some ignoring effects, such as mutual coupling effect, give misleading results in wireless communication systems \cite{Shen10}. Based on different antenna elements, such as dipole, patch and dualpolarized patch antennas, it was demonstrated that the mutual coupling and spatial correlation have practical limit on the spectrum efficiency of multi-user massive MIMO systems \cite{Artiga12}. Considering the spatial correlation and mutual coupling effects on massive MIMO systems, the performance of linear precoders was analyzed for wireless communications systems \cite{Masouros13}. However, in all aforementioned studies, only regular antenna arrays were considered for massive MIMO systems with the mutual coupling effect. {Considering the aesthetic appearance of the commercial buildings, building a platform with a large number of regular antennas on the facade will face confrontations from the building owners. To tackle the challenge of deploying large number of BS antennas, the antennas elements were integrated into the environments, such as a part of the building facade or signage \cite{Boccardi14,Chih-Lin14}. Moreover, the large number of antennas make it very difficult to maintain uniform antenna spacings in these scenarios. As a consequence, these antenna arrays are appropriate to be considered as irregular antenna arrays with nonuniform antenna spacings rather than regular antenna arrays with uniform antenna spacings. For irregular antenna arrays, some studies have been made for conformal antenna arrays where antenna arrays are designed to conform the prescribed shape \cite{Josefsson06,Guo09,Xu08}. Sparse antenna arrays where antenna arrays are configured to decrease the number of antennas but lead to nonuniform antenna spacings and irregular array shapes have also been studied \cite{Chen07,Minvielle11,Hawes12,Cappellen06}. However, these antenna arrays were mainly studied in the field of phased arrays and have never been discussed for massive MIMO communication systems. Motivated by the above gaps, we investigate multi-user massive MIMO wireless communication systems with irregular antenna arrays considering the mutual coupling.} The contributions and novelties of this paper are summarized as follows.

\begin{enumerate}
\item Considering uneven surfaces of antennas deployment regions, a massive MIMO communication system with an irregular antenna array is firstly proposed and formulated. Moreover, the impact of mutual coupling on irregular antenna arrays is evaluated by metrics of the matrix correlation coefficient and ergodic received gain.
\item Based on the results of irregular antenna arrays with mutual coupling, the lower bound of the ergodic achievable rate, average symbol error rate (SER) and average outage probability are derived for multi-user massive MIMO communication systems. Furthermore, asymptotic results are also derived when the number of antennas approaches infinity.
\item Numerical results indicate that there exists a maximum achievable rate for massive MIMO communication systems using irregular antenna arrays. Moreover, the irregular antenna array outperforms the regular antenna array in the achievable rate of massive MIMO communication systems when the number of antennas is larger than or equal to a given threshold.
\end{enumerate}

The remainder of this paper is outlined as follows. Section II describes a system model for massive MIMO communication systems where BS antennas are deployed by an irregular antenna array. In Section III, the impact of mutual coupling on the irregular antenna array is analyzed by the matrix correlation coefficient and ergodic received gain. In section IV, the lower bound of the ergodic achievable rate, average SER and average outage probability are derived for multi-user massive MIMO communication systems using irregular antenna arrays. Considering that the number of antennas approaches infinity, asymptotic results are also obtained. Numerical results and discussions are presented in Section V. Finally, conclusions are drawn in Section VI.

\section{System Model}
\label{sec2}

With the massive MIMO technology emerging in 5G mobile communication systems, hundreds of antennas have to be deployed on the BS tower or the surface of a building. However, surfaces used for deploying massive MIMO antennas are not ideally smooth planes in most of the real scenarios. When massive MIMO antennas have to be deployed on uneven surfaces, spatial distances among adjacent antennas are not expected to be perfectly uniform. In this case, massive MIMO communication systems have to be deployed by irregular antenna arrays. Furthermore, the impact of irregular antenna arrays on massive MIMO communication systems need to be reevaluated when the mutual coupling of irregular antenna arrays is considered. Motivated by above challenges, a single-cell multi-user massive MIMO communication system with an irregular antenna array is illustrated in Fig. 1.

\begin{figure}
\vspace{0.1in}
\centerline{\includegraphics[width=8cm,draft=faulse]{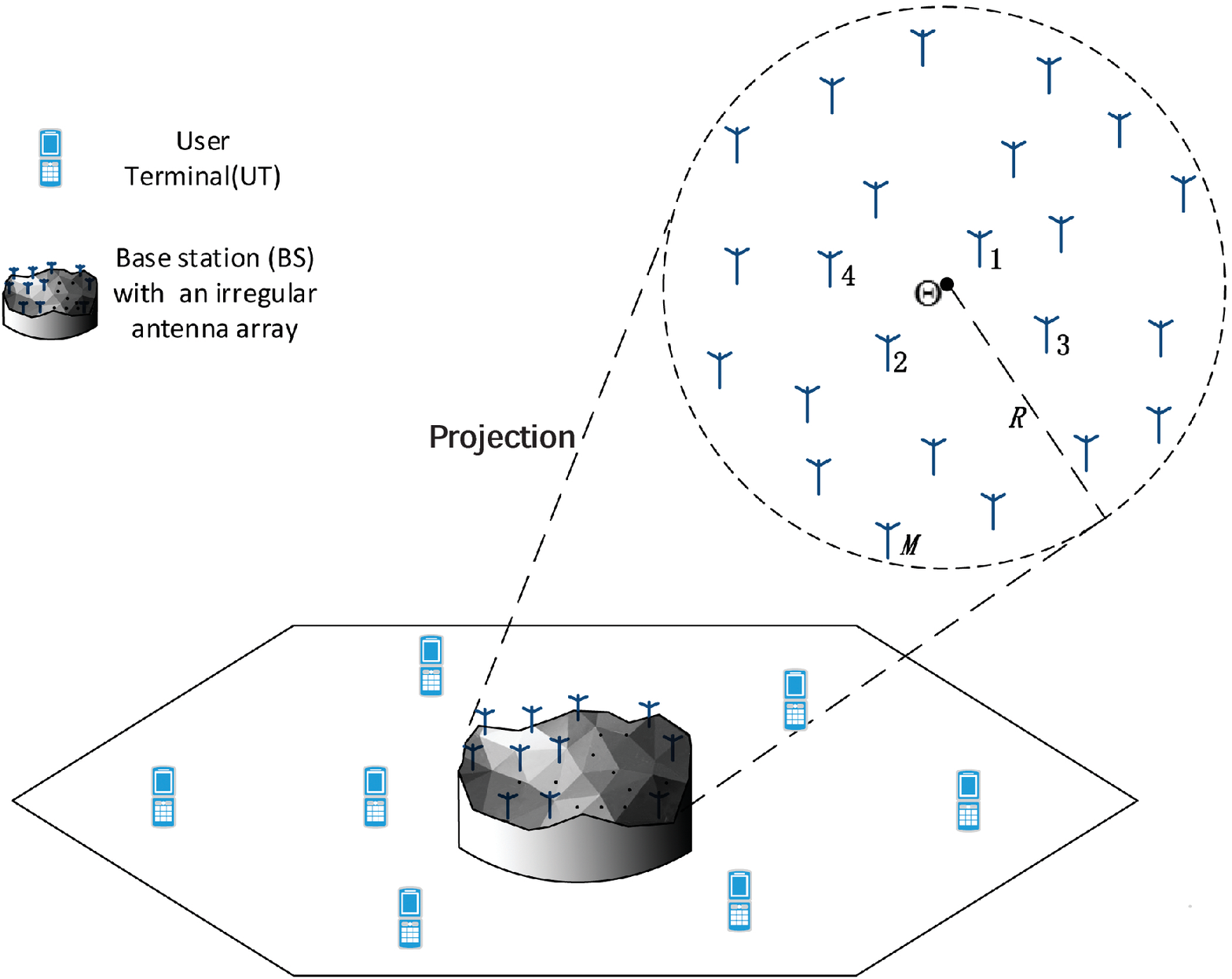}}
{\small Fig. 1. \@ \@ Multi-user massive MIMO communication system with irregular antenna array.}
\end{figure}

In this system model, a BS is located at the cell center and equipped with $M$ antennas which are deployed on an uneven surface. Because of the uneven surface, spatial distances among antennas is no longer regular even when antennas are regularly deployed in a two-dimensional plane. To intuitively illustrate the spatial distances among the irregular antenna array, we project the antenna distances deployed at the uneven surface into a smooth plane meanwhile keeping the spatial distances between each pair of antennas the same as before, as shown in Fig. 1. What needs to be mentioned is that the mutual coupling effect depends on the spatial distance among antennas, which remains the same through this projection. Hence the projecting in Fig. 1 does not affect the mutual coupling effect of irregular antenna array. In the projected plane of Fig. 1, without loss of generality, all antennas are assumed to be covered by a circle centered at $\Theta $ with the radius $R$. The $i{\text{-th}}$ and $j{\text{-th}}$ antennas of the massive MIMO antenna array are denoted as $An{t_i}$ and $An{t_j}$, $i \ne j$, $1 \leqslant i \leqslant M$, $1 \leqslant j \leqslant M$. Spatial distances between the circle center $\Theta $ and locations of the antennas $An{t_i}$ and $An{t_j}$ are denoted as ${d_i}$ and ${d_j}$, respectively. To simplify derivations, all antennas are sorted by the spatial distances between the circle center $\Theta $ and their locations in the circle, i.e., ${d_i} < {d_j}$ if $i < j$. Considering that the number of antennas $M$ and the circle area are fixed in a given scenario, the distribution of the $M$ antennas in the circle is assumed to be governed by a binomial point process (BPP) \cite{Srinivasa10,Handbook10}. {It's notable that the circle area is an assumed area on the smooth projection plane to cover all antennas. The circle area does not depend on the actual shape of the antenna deployment regions. Similarly, other random processes can be used for modeling of the irregular antenna distribution according to the specified requirements.} $K$ active user terminals (UTs) are assumed to be uniformly scattered in a cell and each UT is equipped with an antenna. In this paper, we focus on the uplink transmission of the massive MIMO communication system.

The signal vector received at the BS is expressed as
\[{\mathbf{y}} = \sqrt {SN{R_{UT}}} {\mathbf{Gx}} + {\mathbf{w}},\tag{1}\]
where ${\mathbf{y}} \in {\mathbb{C}^{M \times 1}}$ is the $M \times 1$ received signal vector, ${\mathbf{w}} \in {\mathbb{C}^{M \times 1}}$ is the additive white Gaussian noise (AWGN) with zero mean, i.e., ${\mathbf{w}} \sim \mathcal{C}\mathcal{N}{\text{(0, }}{{\mathbf{I}}_{\mathbf{M}}}{\text{)}}$, ${\mathbf{I}}_{\mathbf{M}}$ is the $M \times M$ unit matrix, ${\mathbf{x}} \in {\mathbb{C}^{K \times 1}}$ is the $K \times 1$  symbol vector transmitted by $K$ UTs. Moreover, the UT transmitting power is normalized as 1. $SN{R_{UT}}$ is the transmitting signal-to-noise ratio (SNR) at the UT and values of $SN{R_{UT}}$ at all UTs are assumed to be equal in this paper. Similar to the finite dimensional physical channel and taking into account the mutual coupling effect between antennas, the $M \times K$ channel matrix ${\mathbf{G}} \in {\mathbb{C}^{M \times K}}$ is extended as \cite{Masouros13,Ngo13J}
\[{\mathbf{G}}{\text{ = }}{\mathbf{CAH}}{{\mathbf{D}}^{1/2}},\tag{2}\]
where ${\mathbf{C}}$ is a mutual coupling matrix, ${\mathbf{A}}$ is an array steering matrix, ${\mathbf{H}}$ is a small scale fading matrix, and ${\mathbf{D}}$ is the large scale fading matrix. {The mutual coupling matrix ${\mathbf{C}}$ and the array steering matrix ${\mathbf{A}}$ are affected by the irregularity of the antenna array. The detailed modeling of these two matrixes will be discussed in the next section.} The large scale fading matrix is a $K \times K$ diagonal matrix and is expressed as
\[{\mathbf{D}} = diag({\beta _1},...,{\beta _k},...,{\beta _K}),\tag{3}\]
the $k{\text{-th}}$ diagonal element of matrix ${\mathbf{D}}$, i.e., ${\beta _k}$, denotes the large scale fading factor in the link of the $k{\text{-th}}$ UT and the BS.

The performance of massive MIMO systems depends critically on the propagation environment, properties of antenna arrays and the number of degrees of freedom offered by the physical channel. The propagation environment offers rich scattering if the number of independent incident directions is large in the angular domain. More precisely, a finite-dimensional channel model is introduced in this paper, where the angular domain is divided into $P$ independent incident directions with $P$ being a large but finite number \cite{Ngo13J}. Each independent incident direction, corresponding to the azimuth angle ${\phi _q}$, ${\phi _q} \in \left[ {0,2\pi } \right],{\text{ }}q = 1, \ldots ,P$, and the elevation angle ${\theta _q}$, ${\theta _q} \in \left[ {{{ - \pi } \mathord{\left/
 {\vphantom {{ - \pi } 2}} \right.
 \kern-\nulldelimiterspace} 2},{\pi  \mathord{\left/
 {\vphantom {\pi  2}} \right.
 \kern-\nulldelimiterspace} 2}} \right]$, is associated with an $M \times 1$ array steering vector ${\mathbf{a}}\left( {{\phi _q},{\theta _q}} \right) \in {\mathbb{C}^{M \times 1}}$. In this case, all independent incident directions are associated with an $M \times P$ array steering matrix ${\mathbf{A}}$ which is given by expression (4)

\[\begin{array}{l}
{\bf{A}} = \left[ {{\bf{a}}\left( {{\phi _1},{\theta _1}} \right),...,{\bf{a}}\left( {{\phi _q},{\theta _q}} \right),...,} \right.\\
\;\;\;\;\;\;\;\left. {{\bf{a}}\left( {{\phi _P},{\theta _P}} \right)} \right] \in {^{M \times P}}
\end{array}. \tag{4}\]

{Without loss of generality, the BS is assumed to be surrounded by a group of scatters and associated with a large but finite number of $P$ independent incident directions \cite{Ngo13J}. Therefore, despite locations of UTs, the uplink signals are scattered by the scatters around the BS and arrive at the BS from the $P$ incident directions.} ${\mathbf{H}} \in {\mathbb{C}^{P \times K}}$ is the $P \times K$ small scale fading matrix and extended as
\[{\mathbf{H}} = \left[ {{{\mathbf{h}}_1},...,{{\mathbf{h}}_k},...,{{\mathbf{h}}_K}} \right] \in {\mathbb{C}^{P \times K}},\tag{5a}\]
\[{{\mathbf{h}}_k} = {\left[ {{h_{k,1}},...,{h_{k,q}},...{h_{k,P}}} \right]^T},\tag{5b}\]
where ${h_{k,q}}$ is the small scale fading factor in the link of the $k{\text{-th}}$ UT and the BS at the $q{\text{-th}}$ independent incident direction, which is governed by a complex Gaussian distribution with zero mean and unit variance, i.e., ${h_{k,q}} \sim \mathcal{C}\mathcal{N}{\text{(0,}}1{\text{)}}$. ${\mathbf{C}} \in {\mathbb{C}^{M \times M}}$ is an $M \times M$ mutual coupling matrix which represents the mutual coupling effect on the irregular antenna array. More specifically, ${\left[ {\mathbf{C}} \right]_{i,j} \ne 0}$, i.e., the element at the $i{\text{-th}}$ row and $j{\text{-th}}$ column of $\mathbf{C}$ denotes the mutual coupling coefficient between the antennas $An{t_i}$ and  $An{t_j}$ in the irregular antenna array. Considering the mutual coupling between antennas, wireless channels in massive MIMO communication systems are assumed to be correlated in this paper.

\section{Irregular Antenna Array with Mutual Coupling}
\label{sec3}

To investigate the impact of the mutual coupling on MIMO communication systems, some studies have been carried out for regular antenna arrays \cite{Shen10,Artiga12,Masouros13,Kildal04,Janaswamy02,Clerckx07,Wallace04}. However, the impact of mutual coupling on massive MIMO communication systems with irregular antenna arrays has been rarely investigated. In irregular antenna arrays, the antenna spacings are no longer uniform and different from those of regular antenna arrays. Because of the irregular and non-uniform antenna spacings, the mutual coupling and spatial correlation of irregular antenna arrays also become different from those of regular antenna arrays. {Considering the impact of the mutual coupling and spatial correlation of antenna arrays on massive MIMO systems, the achievable rate of massive MIMO systems is inevitably affected by the antenna array irregularity.} In this section, the channel correlation model is firstly derived for irregular antenna arrays considering the mutual coupling. Furthermore, the ergodic received gain is defined for evaluating the joint impact of the number of antennas and array size on irregular antenna arrays.

\subsection{Channel correlation model}
 Since each UT is equipped with an antenna and UTs are assumed to be distributed far away from each other, channels of different UTs are assumed to be uncorrelated. In this section, the channel correlation is focused on the side of BS irregular antenna arrays. Based on the channel matrix $\mathbf{G}$ in (2), the channel correlation matrix is expressed by \cite{Chuah02}
\[{\mathbf{\Psi }} = \frac{1}{K}{{\mathbf{D}}^{ - 1}}{\mathbb{E}_{\mathbf{H}}}\left( {{\mathbf{G}}{{\mathbf{G}}^H}} \right) = {\mathbf{CA}}{{\mathbf{A}}^H}{{\mathbf{C}}^H},\tag{6}\]
where ${{\mathbf{D}}^{ - 1}}$ is a normalizing result for the large scale fading, ${\mathbb{E}_{\mathbf{H}}}\left(  \cdot  \right)$ is an expectation operator taken over the small scale fading matrix $\mathbf{H}$ and the superscript $H$ denotes the conjugate transpose of a matrix.
Considering that the distribution of spatial distances among $M$ antennas follows a binomial point process for irregular antenna arrays, the probability density function (PDF) of ${d_i}$ is expressed as  expression (7)\cite{Srinivasa10}

\[{f_{{d_i}}}\left( d \right) = \frac{{2\Gamma \left( {M + 1} \right){{\left( {{R^2} - {d^2}} \right)}^{M - i}}{d^{2i - 1}}}}{{{R^{2M}}\Gamma \left( i \right)\Gamma \left( {M - i + 1} \right)}}, \tag{7}\]

where $\Gamma \left( x \right)$ is a Gamma function. For the antenna $Ant_j$, the PDF of $d_j$ is obtained based on (7) as well. {Note that $d_i$ and $d_j$ are distances from the origin to $Ant_i$ and $Ant_j$ measured on the projection plane, respectively.} The origin, or the circle center $\Theta$, the antenna locations of $Ant_i$ and $Ant_j$ together form a triangle. The triangle's interior angle at its vertex $\Theta$ is denoted as ${\psi _{ij}}$. The distribution of ${\psi _{ij}}$ is assumed to be governed by a uniform distribution in the range of 0 and $\pi$. Based on the law of cosines, the distance between $Ant_i$ and $Ant_j$ is derived as
\[{d_{ij}} = \sqrt {d_i^2 + d_j^2 - 2{d_i}{d_j}\cos {\psi _{ij}}}.\tag{8} \]
When the type of all antennas is assumed to be the dipole antenna, the mutual impedance between $Ant_i$ and $Ant_j$ is expressed as (9a) \cite{Balanis12}
with
\[{Z_{ij}} = \frac{\varepsilon }{{4\pi }}\left[ \begin{array}{l}
2\left( {\gamma  + \ln (\varsigma ) + \int_0^\varsigma  {\frac{{\cos x - 1}}{x}dx} } \right)\\
 - \left( {\gamma  + \ln (\mu ) + \int_0^\mu  {\frac{{\cos x - 1}}{x}dx} } \right)\\
 - \left( {\gamma  + \ln (\rho ) + \int_0^\rho  {\frac{{\cos x - 1}}{x}dx} } \right)\\
 - j\left( \begin{array}{l}
2\int_0^\varsigma  {\frac{{\sin x}}{x}dx}  - \int_0^\mu  {\frac{{\sin x}}{x}dx} \\
 - \int_0^\rho  {\frac{{\sin x}}{x}dx}
\end{array} \right)
\end{array} \right], \tag{9a}\]
\[\varsigma  = \frac{{2\pi {d_{ij}}}}{\lambda },\tag{9b}\]
\[\mu  = \frac{{2\pi }}{\lambda }\left( {\sqrt {{d_{ij}}^2 + {l^2}}  + l} \right),\tag{9c}\]
\[\rho  = \frac{{2\pi }}{\lambda }\left( {\sqrt {{d_{ij}}^2 + {l^2}}  - l} \right),\tag{9d}\]
where $\varepsilon$ is an impedance of free space, $\gamma$ is an Euler-Mascheroni constant, $l$ is an antenna length. Furthermore, the mutual impedance matrix ${{\mathbf{Z}}_{\mathbf{C}}}$ of the irregular antenna array is formed with its element ${Z_{ij}}$, $i \ne j$, located at the $i{\text{-th}}$ line and the $j{\text{-th}}$ column. The $M$ diagonal elements in ${{\mathbf{Z}}_{\mathbf{C}}}$ represent the self-impedances of the $M$ antennas in the irregular antenna array. Considering all configuration parameters of antennas in the irregular antenna array are equal, all diagonal elements in ${{\mathbf{Z}}_{\mathbf{C}}}$ are denoted as ${Z_0}$. Based on the mutual impedance matrix ${{\mathbf{Z}}_{\mathbf{C}}}$, the mutual coupling matrix $\mathbf{C}$ is expressed as \cite{Clerckx07}
\[{\mathbf{C}} = \left( {{Z_0} + {Z_L}} \right){\left( {{Z_L}{{\mathbf{I}}_{\mathbf{M}}} + {{\mathbf{Z}}_{\mathbf{C}}}} \right)^{ - 1}},\tag{10}\]
where $Z_L$ is the load impedance of each antenna. {The $M \times M$ mutual coupling matrix ${\bf{C}}$ denotes the coupling of the received signals caused by the antenna array. Based on the results in \cite{Clerckx07}, the mutual coupling matrix ${\bf{C}}$ can be derived from the load, self and mutual impedances of the antenna nodes as shown in (10).} Considering all configuration parameters of antennas in the irregular antenna array are same, the load impedance of each antenna is assumed to be same.

As for a signal from the $q\text{-th}$ incident direction, the corresponding array steering vector of the irregular antenna array is expressed as ${\mathbf{a}}\left( {{\phi _q},{\theta _q}} \right)$, which is the $q\text{-th}$ column of the steering matrix $\mathbf{A}$. Based on the projected plane in Fig. 1, a polar coordinate system with the origin $\Theta$ is assumed. The antenna with the largest spatial distance $d_M$ to the origin is denoted as $Ant_M$. Moreover, $Ant_M$ is assumed to be located at the polar axis of the polar coordinate system and the corresponding polar coordinate is denoted as $(d_M, 0)$. {In the polar coordinate system, we assume the origin $\Theta$ as the phase response reference point with a zero phase. Considering the incident signal with azimuth angle ${\phi _q}$ and elevation angle ${\theta_q}$, the phase response of the point with the coordinate $\left( {{d_x},{\psi _x}} \right)$ is $\exp \left( { - \frac{{2\pi {d_x}}}{\lambda }\alpha } \right) $ in which $\alpha  = \cos {\psi _x}\cos {\phi _q}\sin {\theta _q} + \sin {\psi _x}\sin {\phi _q}\sin {\theta _q}$ \cite{Balanis12}. Therefore, the phase response of $Ant_M$ with the coordinate $(d_M, 0)$ is given by}
\[{a_M}\left( {{\phi _q},{\theta _q}} \right) = \exp \left( { - j\frac{{2\pi {d_M}}}{\lambda }\sin {\theta _q}\cos {\phi _q}} \right).\tag{11}\]

%$\exp \left[ { - \frac{{2\pi }}{\lambda }\left( {{d_x}\cos {\psi _x}\cos {\phi _q}\sin {\theta _q} + {d_x}\sin {\psi _x}\sin {\phi _q}\sin {\theta _q}} \right)} \right]$

{For $Ant_i$ with the spatial distance $d_i$ in the projected plane of Fig. 1, the corresponding position in the polar coordinate system is denoted as $\left( {{d_i},{\psi }} \right)$. Similarly, given the polar coordinate of $Ant_i$, its phase response with the origin $\Theta$ as the phase response reference is derived by expression (12)}

\[\begin{array}{*{20}{l}}
\begin{array}{l}
{a_i}\left( {{\phi _q},{\theta _q}} \right) = \exp \left[ { - \frac{{2\pi }}{\lambda }\left( {{d_i}\cos {\psi _i}\cos {\phi _q}\sin {\theta _q} + } \right.} \right.\\
\left. {\left. {\;\;\;\;\;\;\;\;\;\;\;\;\;\;\;\;\;\;\;\;\;\;\;\;\;{d_i}\sin {\psi _i}\sin {\phi _q}\sin {\theta _q}} \right)} \right]
\end{array}\\
\begin{array}{l}
\;\;\;\;\;\;\;\;\;\;\;\;\;\;\;\; = \exp \left[ { - \frac{{2\pi {d_i}\sin {\theta _q}}}{\lambda }\left( {\cos {\psi _i}\cos {\phi _q} + } \right.} \right.\\
\;\;\;\;\;\;\;\;\;\;\;\;\;\;\;\;\;\;\;\;\;\;\;\;\;\left. {\left. {\sin {\psi _i}\sin {\phi _q}} \right)} \right]\;
\end{array}\\
{\;\;\;\;\;\;\;\;\;\;\;\;\;\;\;\;\;\; = \exp \left[ { - j\frac{{2\pi {d_i}}}{\lambda }\sin {\theta _q}\cos \left( {{\phi _q} - {\psi _i}} \right)} \right].}
\end{array}. \tag{12}\]

Based on (11) and (12), all $M$ elements of ${\mathbf{a}}\left( {{\phi _q},{\theta _q}} \right)$ can be derived. Furthermore, all elements of array steering matrix $\mathbf{A}$ are derived based on (4), (11) and (12).

When derivation results of the mutual coupling matrix $\mathbf{C}$ and array steering matrix $\mathbf{A}$ are substituted into (6), the channel correlation matrix of irregular antenna array is derived. As for MIMO systems, the strength of the channel correlation greatly affects the performance of wireless communications \cite{Kildal04}, \cite{Alfano04}. Moreover, the strength of channel correlation depends on the channel correlation matrix's off-diagonal elements in conventional MIMO systems with regular antenna arrays \cite{Chuah02}. For regular antenna arrays, the channel correlation matrix is a Toeplitz matrix, whose off-diagonal elements get smaller values when the elements' positions get farther from the matrix's diagonal. In general, the magnitude of one of the off-diagonal elements which is the closest to the diagonal of the Toeplitz matrix is selected to represent the strength of the channel correlation for regular antenna arrays. However, spatial distances among antennas are not identical for irregular antenna arrays. In this case, values of the channel correlation matrix's off-diagonal elements do not get smaller when their positions get farther from the matrix's diagonal. Therefore, it is impossible to evaluate the strength of channel correlation with only one element of the channel correlation matrix in irregular antenna arrays. To find a representation of the channel correlation strength for irregular antenna arrays, the matrix correlation coefficient $\eta$ is defined as the ratio of the sum of squared off-diagonal elements to the sum of squared diagonal elements in the channel correlation matrix, which is expressed by
\[\eta  = \frac{{\text{Tr}\left[ {{\mathbf{\Psi }}{{\mathbf{\Psi }}^H}} \right]}}{{\text{Tr}\left[ {{\mathbf{\Psi }} \circ {\mathbf{\Psi }}} \right]}} - 1,\tag{13}\]
where $\text{Tr}$ is the trace operator for matrixes, $ \circ $ is the Hardmard product operator for matrixes. Based on the definition in (13), the value of the matrix correlation coefficient $\eta$ increases with the increase of the strength of channel correlation in irregular antenna arrays.

 \subsection{Ergodic Received Gain}
When the deployment area is fixed for massive MIMO antenna arrays, the number of antennas is inversely proportional to the antenna spatial distance in regular antenna arrays. The received SNR of MIMO antenna arrays rises with the increasing of the received diversity when the number of antennas is increased. However, the antenna spatial distance decreases with the increase of the number of antennas in a fixed deployment area, which leads to the received SNR to decrease with the strengthening of the channel correlation \cite{Masouros13}. Hence, the number of antennas and antenna spatial distance are contradictory parameters for the received SNR of massive MIMO communication systems in a fixed deployment area. When the deployment area of the irregular antenna array is fixed, it is also a critical challenge to evaluate the impact of the number of antennas and antenna spatial distance on the received SNR of massive MIMO communication systems with irregular antenna arrays. To easily investigate the impact of the varying number of antennas and antenna spatial distance to the received SNR, the ergodic received gain of irregular antenna arrays is defined as

\[\mathbb{G}\left( {M,R} \right) = \mathbb{E}\left( {SN{R_{BS}} - SNR_{BS}^{\min }} \right),\tag{14}\]
where $SNR_{BS}^{\min }$ refers to the received SNR when the number of antennas is configured as its available minimum value ${M_{\min }}$ and the circle radius in Fig. 1 is configured as its available minimum value ${R_{\min }}$. ${M_{\min }}$ and ${R_{\min }}$ are configured as constants in this paper and $SNR_{BS}^{\min }$ can be viewed as a reference point for the received SNR.  $SN{R_{BS}}$ refers to the received SNR with the number of antennas $M$ and the circle radius $R$. {Comparing with $SNR_{BS}$, the ergodic received gain in (14) can directly reflect the increment of the received SNR when the number of antennas $M$ and the circle radius $R$ keep increasing from their minimum values.}  Moreover, the ergodic received gain also reflects how the achievable rate changes with the varying number of antennas and antenna spatial distances since the achievable rate monotonously increases with the increase of received SNR in wireless communication systems.

To simplify performance analysis of the ergodic received gain in irregular antenna arrays, only one UT is assumed in this case and the BS has perfect channel state information. Moreover, the maximum ratio combining (MRC) detector scheme is adopted in the BS. In this case, the BS received signal is expressed as
%\[\begin{gathered}
%{\tilde y_{1,MRC}} = \sqrt {SN{R_{UT}}} {\beta _1}{\mathbf{h}}_1^H{{\mathbf{A}}^H}{{\mathbf{C}}^H}{\mathbf{CA}}{{\mathbf{h}}_1}x{\kern 1pt}\\
%+ \beta _1^{1/2}{\mathbf{h}}_1^H{{\mathbf{A}}^H}{{\mathbf{C}}^H}{w_1}
%\end{gathered},\tag{15}\]
\[\begin{gathered}
  {{\tilde y}_{1,MRC}} = \sqrt {SN{R_{UT}}} {\beta _1}{\mathbf{h}}_1^H{{\mathbf{A}}^H}{{\mathbf{C}}^H}{\mathbf{CA}}{{\mathbf{h}}_1}x \hfill \\
  {\kern 1pt} {\kern 1pt} {\kern 1pt} {\kern 1pt} {\kern 1pt} {\kern 1pt} {\kern 1pt} {\kern 1pt} {\kern 1pt} {\kern 1pt} {\kern 1pt} {\kern 1pt} {\kern 1pt} {\kern 1pt} {\kern 1pt} {\kern 1pt} {\kern 1pt} {\kern 1pt} {\kern 1pt} {\kern 1pt} {\kern 1pt} {\kern 1pt} {\kern 1pt} {\kern 1pt} {\kern 1pt} {\kern 1pt} {\kern 1pt} {\kern 1pt} {\kern 1pt} {\kern 1pt} {\kern 1pt} {\kern 1pt} {\kern 1pt}  + \beta _1^{1/2}{\mathbf{h}}_1^H{{\mathbf{A}}^H}{{\mathbf{C}}^H}{w_1} \hfill \\
\end{gathered},\tag{15} \]

where $x$ is the symbol transmitted by the UT, ${\omega _1}$ is the AWGN with zero mean over wireless channels. {$SN{R_{UT}}$ is the transmit SNR at the UT and equals to the transmit power of the UT.} Furthermore, the received SNR at BS is given by
\[\begin{gathered}
  SN{R_{1,BS,MRC}} = \frac{{{{\left| {\sqrt {SN{R_{UT}}} {\beta _1}{\mathbf{h}}_1^H{{\mathbf{A}}^H}{{\mathbf{C}}^H}{\mathbf{CA}}{{\mathbf{h}}_1}x} \right|}^2}}}{{{{\left| {\beta _1^{1/2}{\mathbf{h}}_1^H{{\mathbf{A}}^H}{{\mathbf{C}}^H}{w_1}} \right|}^2}}} \hfill \\
  \;\;\;\;\;\;\;\;\;\;\;\;\;\;\;\;\;\;\;\;\;\;\;\; = SN{R_{UT}}{\beta _1}{\mathbf{h}}_1^H{{\mathbf{A}}^H}{{\mathbf{C}}^H}{\mathbf{CA}}{{\mathbf{h}}_1} \hfill \\
\end{gathered}.\tag{16} \]
{For the single UT scenario, the minimum received SNR with the BS irregular antenna array, which is denoted as $SNR_{1,BS,MRC}^{\min }$, is expressed by}
\[SNR_{1,BS,MRC}^{\min } = SN{R_{UT}}{\beta _1}{\mathbf{h}}_1^H{{\mathbf{\hat A}}^H}{{\mathbf{\hat C}}^H}{\mathbf{\hat C\hat A}}{{\mathbf{h}}_1},\tag{17}\]
where ${\mathbf{\hat C}}$ and ${\mathbf{\hat A}}$ are the mutual coupling matrix and array steering matrix with the minimum number of antennas ${M_{\min }}$ and the minimum circle radius ${R_{\min }}$, respectively. Based on the definition of ergodic received gain in (14), the ergodic received gain with the single UT is derived by
\[\begin{gathered}
  \mathbb{G}\left( {M,R} \right) = \mathbb{E}\left( {SN{R_{1,BS,MRC}} - SNR_{1,BS,MRC}^{\min }} \right) \hfill \\
  \;\;\;\;\;\;\;\;\;\;\;\;\;\;\; = SN{R_{UT}}{\beta _1}\left[ {\mathbb{E}\left( {{\xi _1}} \right) - \mathbb{E}\left( {{\xi _{\min }}} \right)} \right] \hfill \\
\end{gathered},\tag{18a} \]
with
\[{\xi _1} = {\mathbf{h}}_1^H{{\mathbf{A}}^H}{{\mathbf{C}}^H}{\mathbf{CA}}{{\mathbf{h}}_1},\tag{18b}\]
\[{\xi _{\min }} = {\mathbf{h}}_1^H{{\mathbf{\hat A}}^H}{{\mathbf{\hat C}}^H}{\mathbf{\hat C\hat A}}{{\mathbf{h}}_1}.\tag{18c}\]

The conventional antenna arrays have been investigated based on the regular antenna distance \cite{Handbook10}, \cite{Chuah02,Matthaiou03,Kiessling03,Alrabadi13}. However, it is difficulty to derive an analytical expression for the ergodic received gain when the random distributed antenna spatial distances are presented at irregular antenna arrays. As a matter of fact, the channel correlation matrix ${\mathbf{\Psi }} = {\mathbf{CA}}{{\mathbf{A}}^H}{{\mathbf{C}}^H}$ is modeled based on the random distributed antenna spatial distances. Comparing with the fast varying small scale fading matrix $\mathbf{H}$, $\mathbf{\Psi }$ can be viewed as a deterministic matrix just like the large scale fading matrix $\mathbf{D}$. Therefore, to simplify the derivation in (18), the expectation in (18a) is assumed to be taken over the small scale fading vector, i.e., ${{\mathbf{h}}_1}$. In addition, all eigenvalues ${\tau _p},{\text{ }}1 \leqslant p \leqslant M$, of channel correlation matrix ${\mathbf{\Psi }} = {\mathbf{CA}}{{\mathbf{A}}^H}{{\mathbf{C}}^H}$ are assumed to be known. Therefore, the following proposition is derived.

${\textbf{\emph{Proposition 1}}}$: For the single UT scenario, the BS has the perfect channel state information and adopts the MRC detector scheme, the ergodic received gain at massive MIMO communication systems with irregular antenna arrays is derived by expression (19a)

\[\begin{gathered}
  \mathbb{G}\left( {M,R} \right) = SN{R_{UT}}{\beta _1}\left[ {\frac{{\det \left( {{{\mathbf{B}}_{M,1}}} \right)}}{{\prod\limits_{i < j}^M {\left( {{\tau _j} - {\tau _i}} \right)} }} \times } \right. \hfill \\
  \;\;\;\;\;\;\;\;\left( {\tau _M^M - \sum\limits_{p = 1}^{M - 1} {\sum\limits_{q = 1}^{M - 1} {{{\left[ {{\mathbf{B}}_{M,1}^{ - 1}} \right]}_{q,p}}\tau _M^{q - 1}\tau _p^M} } } \right) \hfill \\
  \;\;\;\;\;\;\;\; - \frac{{\det \left( {{{{\mathbf{\hat B}}}_{{M_{\min }},1}}} \right)}}{{\prod\limits_{i < j}^{{M_{\min }}} {\left( {{\tau _j} - {\tau _i}} \right)} }}\left( {\hat \tau _{{M_{\min }}}^{{M_{\min }}} - } \right. \hfill \\
  \left. {\left. {\;\;\;\;\;\sum\limits_{p = 1}^{{M_{\min }} - 1} {\sum\limits_{q = 1}^{{M_{\min }} - 1} {{{\left[ {{\mathbf{\hat B}}_{{M_{\min }},1}^{ - 1}} \right]}_{q,p}}\hat \tau _{{M_{\min }}}^{q - 1}\hat \tau _p^{{M_{\min }}}} } } \right)} \right] \hfill \\
\end{gathered}, \tag{19a}\]

with
\[{{\mathbf{B}}_{M,1}} = \left[ {\begin{array}{*{20}{c}}
  1&{{\tau _1}}& \cdots  \\
   \vdots & \vdots & \ddots  \\
  1&{{\tau _{M - 1}}}& \ldots
\end{array}\;\;\;\begin{array}{*{20}{c}}
  {\tau _1^{M - 2}} \\
   \vdots  \\
  {\tau _{M - 1}^{M - 2}}
\end{array}} \right],\tag{19b}\]
\[{{\mathbf{\hat B}}_{{M_{\min }},1}} = \left[ {\begin{array}{*{20}{c}}
  1&{{{\hat \tau }_1}}& \cdots  \\
   \vdots & \vdots & \ddots  \\
  1&{{{\hat \tau }_{{M_{\min }} - 1}}}& \ldots
\end{array}\;\;\;\begin{array}{*{20}{c}}
  {\hat \tau _1^{{M_{\min }} - 2}} \\
   \vdots  \\
  {\hat \tau _{{M_{\min }} - 1}^{{M_{\min }} - 2}}
\end{array}} \right],\tag{19c}\]
where ${\hat \tau _p},{\text{ }}1 \leqslant p \leqslant {M_{\min }}$ is the eigenvalue of channel correlation matrix ${\mathbf{\hat \Psi }} = {\mathbf{\hat C\hat A}}{{\mathbf{\hat A}}^H}{{\mathbf{\hat C}}^H}$.

${\textbf{\emph{Proof}}}$: Based on the BS configuration and the single UT scenario, the ergodic received gain is expressed in (18a), (18b) and (18c). When all eigenvalues ${\tau _p},{\text{ }}1 \leqslant p \leqslant M$, of channel correlation matrix ${\mathbf{\Psi }} = {\mathbf{CA}}{{\mathbf{A}}^H}{{\mathbf{C}}^H}$ are assumed to be known, the conditional PDF of ${\xi _1}$ is derived by expression (20) \cite{Alfano04}

\[\begin{array}{l}
{f_{{\xi _1}}}\left( {x|{\tau _1},...,{\tau _M}} \right) = \frac{{\det \left( {{{\bf{B}}_{M,1}}} \right)}}{{\prod\limits_{i < j}^M {\left( {{\tau _j} - {\tau _i}} \right)} }}\left( {\tau _M^{M - 2}{e^{ - x/{\tau _M}}}} \right.\\
\left. {\;\;\;\;\;\;\;\;\; - \sum\limits_{p = 1}^{M - 1} {\sum\limits_{q = 1}^{M - 1} {{{\left[ {{\bf{B}}_{M,1}^{ - 1}} \right]}_{q,p}}\tau _M^{q - 1}\tau _p^{M - 2}{e^{ - x/{\tau _p}}}} } } \right)
\end{array}, \tag{20}\]

Furthermore, the term of ${\mathbb{E}_{{{\mathbf{h}}_1}}}\left( {{\xi _1}} \right)$ is derived by expression (21),
where (a) is obtained because ${\xi _1} = {\mathbf{h}}_1^H{{\mathbf{A}}^H}{{\mathbf{C}}^H}{\mathbf{CA}}{{\mathbf{h}}_1} \geqslant 0$.

\[\begin{gathered}
  {\mathbb{E}_{{{\mathbf{h}}_1}}}\left( {{\xi _1}} \right) = \int\limits_{ - \infty }^{ + \infty } {x{f_{{\xi _1}}}\left( {x|{\tau _1},...,{\tau _M}} \right)} {\text{d}}x \hfill \\
  \mathop  = \limits^{(a)} \frac{{\det \left( {{{\mathbf{B}}_{M,1}}} \right)}}{{\prod\limits_{i < j}^M {\left( {{\tau _j} - {\tau _i}} \right)} }}\left( {\tau _M^{M - 2}\int\limits_0^{ + \infty } {x{e^{ - x/{\tau _M}}}} {\text{d}}x} \right. \hfill \\
  \left. {\;\;\; - \sum\limits_{p = 1}^{M - 1} {\sum\limits_{q = 1}^{M - 1} {{{\left[ {{\mathbf{B}}_{M,1}^{ - 1}} \right]}_{q,p}}\tau _M^{q - 1}\tau _p^{M - 2}\int\limits_0^{ + \infty } {x{e^{ - x/{\tau _M}}}} {\text{d}}x} } } \right) \hfill \\
   = \frac{{\det \left( {{{\mathbf{B}}_{M,1}}} \right)}}{{\prod\limits_{i < j}^M {\left( {{\tau _j} - {\tau _i}} \right)} }}\left( {\tau _M^M - } \right.\left( {\sum\limits_{p = 1}^{M - 1} {\sum\limits_{q = 1}^{M - 1} {{{\left[ {{\mathbf{B}}_{M,1}^{ - 1}} \right]}_{q,p}}\tau _M^{q - 1}\tau _p^M} } } \right) \hfill \\
\end{gathered}, \tag{21} \]

Similarly, the term of ${\mathbb{E}_{{{\mathbf{h}}_1}}}\left( {{\xi _{\min }}} \right)$ is derived by expression (22),

\[\begin{gathered}
  {\mathbb{E}_{{{\mathbf{h}}_1}}}\left( {{\xi _{\min }}} \right) = \frac{{\det \left( {{{{\mathbf{\hat B}}}_{{M_{\min }},1}}} \right)}}{{\prod\limits_{i < j}^{{M_{\min }}} {\left( {{\tau _j} - {\tau _i}} \right)} }}\left( {\hat \tau _{{M_{\min }}}^{{M_{\min }}}} \right. \hfill \\
  \left. {\;\;\;\;\;\;\;\; - \sum\limits_{p = 1}^{{M_{\min }} - 1} {\sum\limits_{q = 1}^{{M_{\min }} - 1} {{{\left[ {{\mathbf{\hat B}}_{{M_{\min }},1}^{ - 1}} \right]}_{q,p}}\hat \tau _{{M_{\min }}}^{q - 1}\hat \tau _p^{{M_{\min }}}} } } \right) \hfill \\
\end{gathered}, \tag{22} \]

Substitute (21) and (22) into (18a), the expression of (19) is obtained. {Meanwhile, it's worth mentioning that besides equation (20) from \cite{Alfano04}, Theorem 1 in \cite{Jin10} also can be used to obtain the PDF of ${\xi _1}$ and derive the closed form result of the ergodic received gain.}

\begin{figure}
\vspace{0.1in}
\centerline{\includegraphics[width=9.5cm,draft=false]{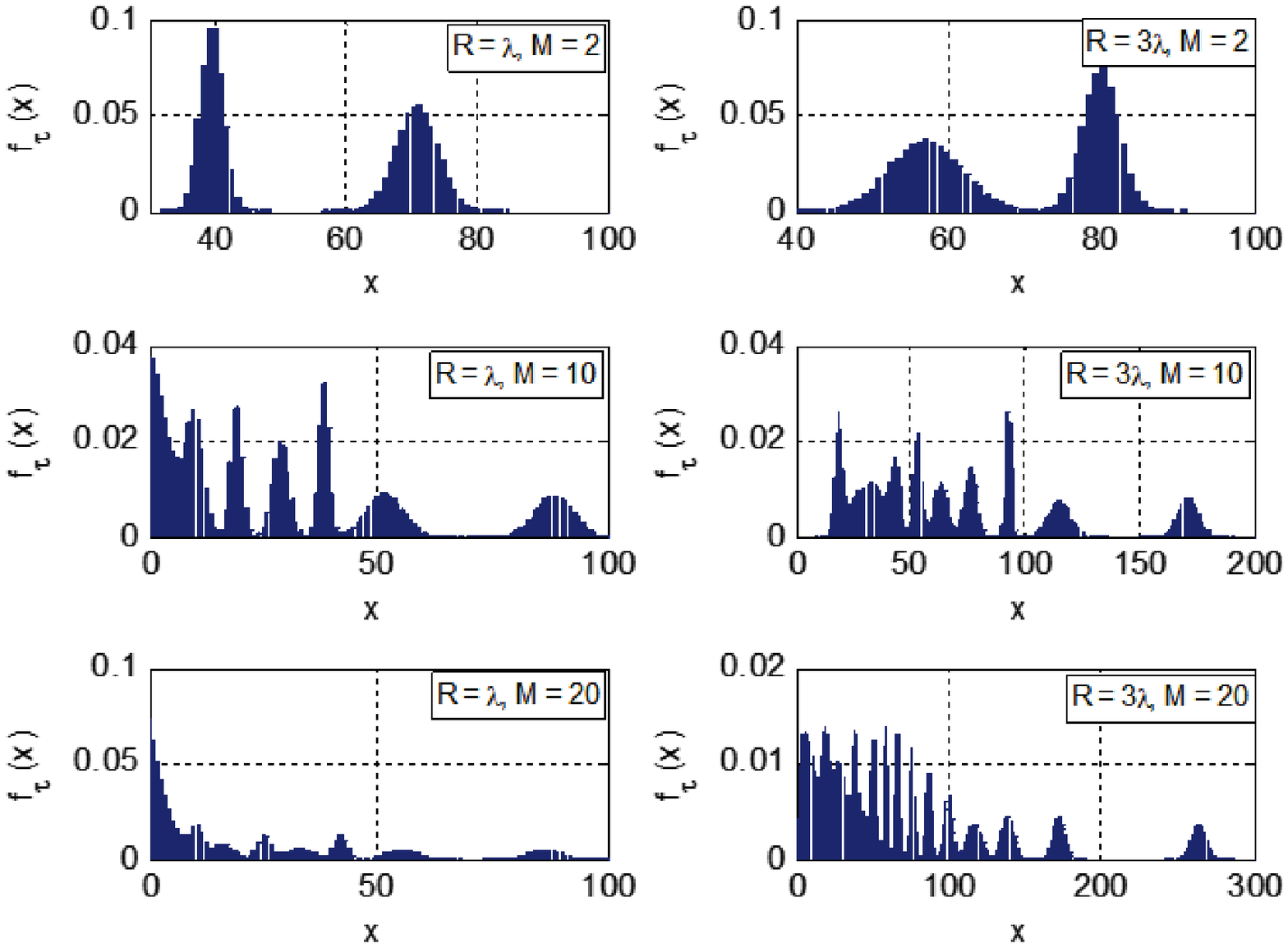}}
{\small Fig. 2. \@ \@ Empirical probability distribution of the eigenvalues of the channel correlation matrix ${\mathbf{\Psi }}$.}
\end{figure}

\subsection{Numerical analysis}
Based on definitions of the matrix correlation coefficient and ergodic received gain, some performance evaluations can be numerically analyzed in detail. In the following analysis, default parameters in Fig. 1 are configured: antennas are assumed to be dipole antennas \cite{Balanis12}, the load impedance of every antenna is ${Z_L} = 50$ Ohms, the self-impedance of every antenna is 50 Ohms \cite{Shen10}, the carry frequency used for wireless communications is 2.5 GHz, the corresponding wavelength is $\lambda  = 0.12$ meter (m), the number of independent incident directions in propagation environments is configured as $P = 100$ \cite{Ngo13J}.

In the performance analysis of conventional regular antenna arrays, the channel correlation matrix is a Toeplitz matrix where the eigenvalues of Toeplitz matrix converge to some limited distributions \cite{Chuah02}. However, the channel correlation matrix of irregular antenna array is not a Toeplitz matrix. In this case, the eigenvalues of the channel correlation matrix is analytically intractable. Therefore, we try to obtain the eigenvalues for irregular antenna arrays empirically by numerical simulations. When the circle radius $R$ is configured as $\lambda$ and $3\lambda$, Fig. 2 shows the empirical distribution of the eigenvalues of the channel correlation matrix ${\mathbf{\Psi }}$ with the number of antennas $M=2$, $M=10$ and $M=20$, respectively. Furthermore, we try to select suitable distributions to best match the simulation results in Fig. 2. The sum of normal distributions appears to agree well with the simulation results. To simplify the matching results, each eigenvalue of the channel correlation matrix is approximated as the expectation of each normal distribution. For example, for the case with $R = \lambda$ and $M = 2$, the eigenvalues are approximated as 38.4 and 72.3. When the approximated eigenvalues are substituted into (19), the ergodic received gain is obtained for the irregular antenna array. {It's worth mentioning that the BPP is used to model a specific antenna deployment scenario, the eigenvalues' distributions in Fig. 2 are applicable for the specified scenarios with the BPP antenna distribution.}

\begin{figure}
\vspace{0.1in}
\centerline{\includegraphics[width=9.5cm,draft=false]{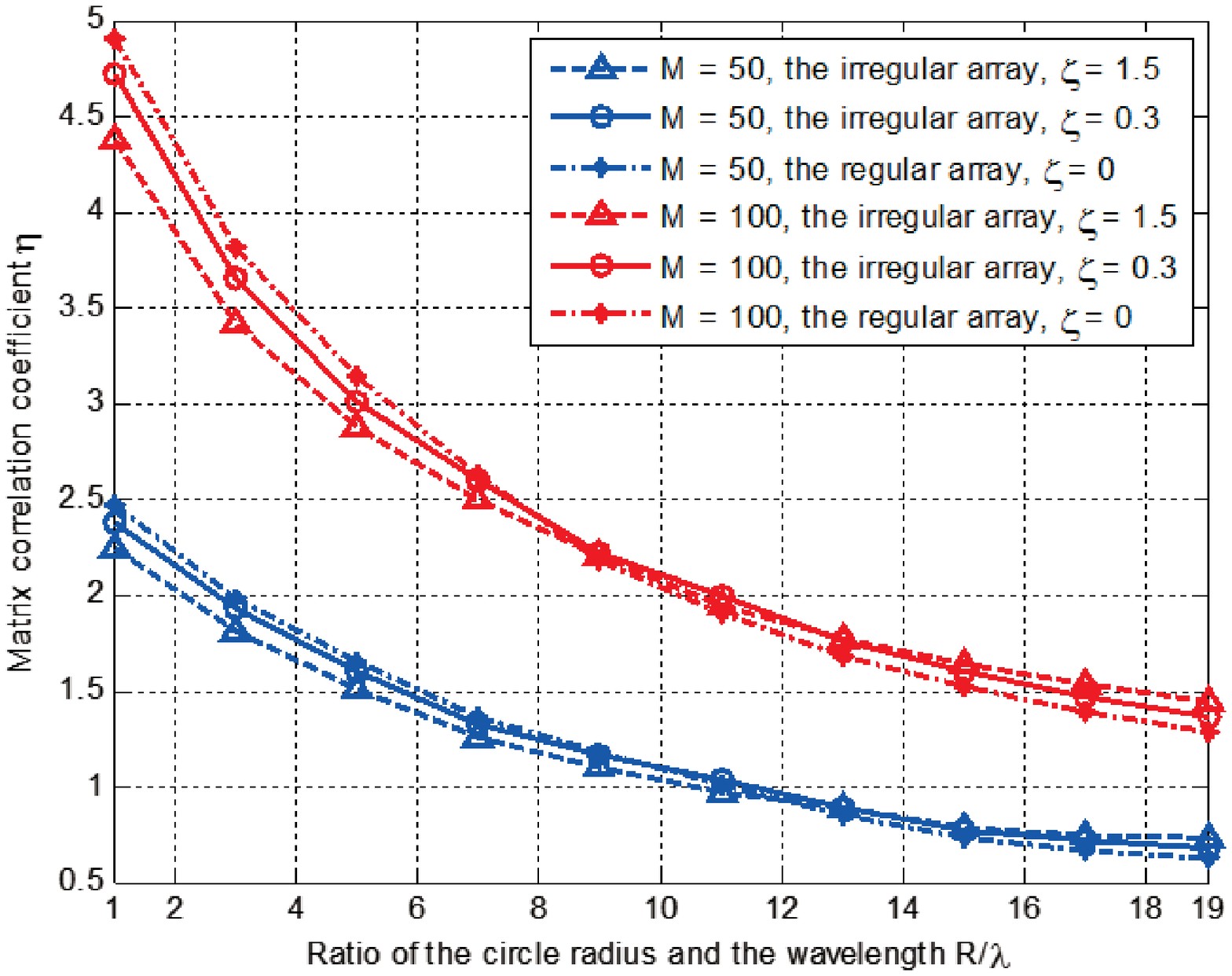}}
{\small Fig. 3. \@ \@ Matrix correlation coefficient with respective to the ratio of the circle radius $R$
and the wavelength $\lambda$, the number of antennas $M$ and the irregularity coefficient $\zeta$.}
\end{figure}

To illustrate how the irregularity of the antenna array affects the performance of massive MIMO systems, the irregularity coefficient $\zeta $ is defined here. As introduced in \cite{Doswell97}, the parameter $\zeta $  is used to model the degree of irregularity of the points' distribution on a plane. In this paper, we use $\zeta $  to describe the irregularity of the antenna distribution on the projection plane. According to the definition in [39], $\zeta $ is the magnitude of the gradient of the sum of the weights, i.e. $\zeta  = \left| {\nabla \left( {\sum\limits_{i = 1}^M {{w_i}} } \right)} \right|$, {in which the weight $w_i$ is calculated as ${w_i} = {e^{ - {{\left( {{{{d_i}} \mathord{\left/
 {\vphantom {{{d_i}} {1.3}}} \right.
 \kern-\nulldelimiterspace} {1.3}}} \right)}^2}}}$, where $d_i$ is the distance from the origin to the antenna $Ant_i$ and the constant 1.3 is an empirical parameter \cite{Doswell97}.} $\zeta$ is a nonnegative value. For a regular antenna array with identical antenna spacing, the value of $\zeta$ is 0. For an irregular antenna array whose antenna distribution follows the BPP distribution, the value of $\zeta $ is approximately 1.5. Moreover, the larger value of $\zeta $ corresponds to the higher level of irregularity in the antenna array. In Fig. 3, the matrix correlation coefficient with respective to the ratio of the circle radius and the wavelength, the number of antennas, and the irregularity coefficient are illustrated. It can be seen in Fig. 3 that the matrix correlation coefficient decreases with the increase of the ratio of the circle radius and the wavelength when the number of antennas and the irregular coefficient of the antenna array are fixed. When the ratio of the circle radius and the wavelength is fixed, the larger number of antennas corresponds to the larger value of the matrix correlation coefficient. Furthermore, when the ratio of the circle radius and the wavelength is small, the smaller value of the irregularity coefficient corresponds to the larger correlation coefficient of the array. But with the increasing of the ratio of the circle radius and the wavelength, the curves with different irregularity coefficients get crossed. When the ratio of the circle radius and the wavelength is large, the smaller value of the irregularity coefficient corresponds to the smaller correlation coefficient of the array. This phenomenon shows the irregularity of the antenna array helps to decrease the correlation of antenna arrays when the size of the array, i.e. the ratio of the circle radius and the wavelength, is less than a given threshold, but increase the correlation when the size of the array is larger than or equal to a given threshold. When the array size is less than the cross point, the expectation of the average antenna spacing in the irregular array is larger than the average antenna spacing in the regular array. But when the array size is larger than the cross point, the expectation of the average antenna spacing in the irregular array is less than that in the regular array. Because the smaller average antenna spacing corresponds to the higher antenna correlation, the curves of regular antenna arrays and irregular antenna arrays cross each other when the array size increases a given threshold.

\begin{figure}
\vspace{0.1in}
\centerline{\includegraphics[width=9.5cm,draft=false]{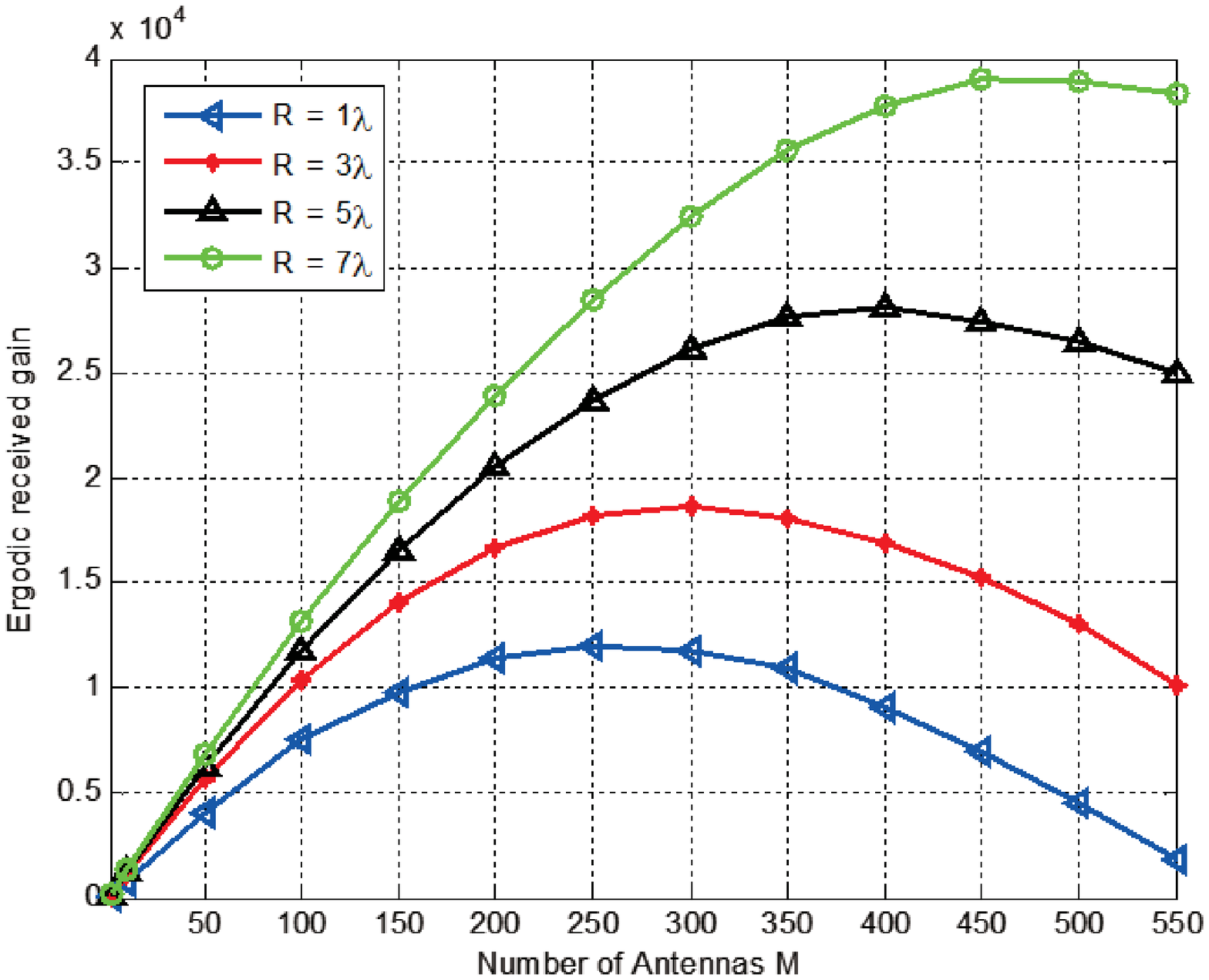}}
{\small
Fig. 4. \@ \@ Ergodic received gain with respect to the number of antennas and the circle radius.}
\end{figure}

The impact of the number of antennas and the circle radius on the ergodic received gain of irregular antenna arrays with $\zeta=1.5$ has been investigated in Fig. 4. It should be mentioned that in the rest of this paper, if there isn't any specific notifications, the term ``{\em irregular antenna}" refers to the irregular antenna array with irregularity coefficient $\zeta=1.5$. Considering the single UT scenario, default parameters are configured as follows: the available minimum radius is ${R_{\min }} = \lambda $, the available minimum number of antennas is ${M_{\min }} = 2 $, the large scale fading factor is ${\beta _1} = {{z \mathord{\left/
 {\vphantom {z {\left( {{{{l_1}} \mathord{\left/
 {\vphantom {{{l_1}} {{l_{{\text{resist}}}}}}} \right.
 \kern-\nulldelimiterspace} {{l_{{\text{resist}}}}}}} \right)}}} \right.
 \kern-\nulldelimiterspace} {\left( {{{{l_1}} \mathord{\left/
 {\vphantom {{{l_1}} {{l_{{\text{resist}}}}}}} \right.
 \kern-\nulldelimiterspace} {{l_{{\text{resist}}}}}}} \right)}}^v}$, where $z$ is the random variable of the log-normal distribution with standard variance ${\sigma _{{\text{shadow}}}} = 8$ dB, the distance between the UT and the BS is ${l_1} = 100$ m, the protect distance between the UT and the BS is ${l_{{\text{resist}}}} = 10$ m, and the path loss coefficient is $v = 3.8$. When the number of antennas is fixed, the ergodic received gain of irregular antenna array increases with the increase of the circle radius. {Numerical results indicate that the ergodic received gain firstly increases with the increase of the number of antennas, but then decreases when the number of antennas at irregular antenna arrays exceeds a given threshold. These results imply that there exists a maximal value for the ergodic received gain of irregular antenna arrays. The maximal ergodic received gain values are $1.2 \times {10^4}$, $1.86 \times {10^4}$, $2.8 \times {10^4}$ and $3.9 \times {10^4}$, corresponding to the numbers of antennas as 250, 300, 400 and 450, respectively.}

\section{Multi-user Massive MIMO Communication Systems}
Based on the ergodic received gain of irregular antenna arrays, the lower bound of the ergodic achievable rate, the average SER and the average outage probability are presented for multi-user massive MIMO communication systems with irregular antenna arrays in this section. {It's notable that in this section we firstly consider the massive MIMO system with hundreds of BS antennas and use a finite-dimensional method to investigate its performance. Then, the case with an infinite number of BS antennas is considered and asymptotic results are obtained for massive MIMO system performance metrics.}

\subsection{Achievable rate}
Assume that the zero-forcing detector is adopted at the BS to cancel the inter-user interference in the cell in Fig. 1. The BS detecting matrix is denoted as ${\mathbf{F}} = {\mathbf{G}}{\left( {{{\mathbf{G}}^H}{\mathbf{G}}} \right)^{ - 1}} \in {\mathbb{C}^{M \times K}}$. Therefore, the received signal at the BS is expressed as
\[\begin{gathered}
  {\mathbf{\tilde y}} = {{\mathbf{F}}^H}{\mathbf{y}}{\kern 1pt} \; \hfill \\
  \;\;\; = \sqrt {SN{R_{UT}}} {{\mathbf{F}}^H}{\mathbf{Gx}}{\kern 1pt}  + {{\mathbf{F}}^H}{\mathbf{w}} \hfill \\
  \;\;\; = \sqrt {SN{R_{UT}}} {\mathbf{x}}{\kern 1pt}  + {{\mathbf{F}}^H}{\mathbf{w}} \hfill \\
\end{gathered}. \tag{23}\]

Considering the BS received signals transmitted from $K$ active UTs, ${\mathbf{\tilde y}} = \left[ {{{\tilde y}_1}, \cdots ,{{\tilde y}_k}, \cdots ,{{\tilde y}_K}} \right] \in {\mathbb{C}^{K \times 1}}$ is a $K \times 1$ vector. The BS received signal transmitted from the $k{\text{-th}}$ UT is expressed as
\[{\tilde y_k} = \sqrt {SN{R_{UT}}} {x_k}{\kern 1pt}  + {\mathbf{F}}_k^H{\mathbf{w}}.\tag{24}\]
Furthermore, the BS received SNR over the link of the $k{\text{-th}}$ UT is expressed by
\[SN{R_{k,BS,ZF}} = \frac{{SN{R_{UT}}}}{{{{\left\| {{\mathbf{F}}_k^H} \right\|}^2}}} = \frac{{SN{R_{UT}}}}{{{{\left[ {{{\left( {{{\mathbf{G}}^H}{\mathbf{G}}} \right)}^{ - 1}}} \right]}_{kk}}}}.\tag{25}\]
As a consequence, the achievable rate for the $k{\text{-th}}$ UT is derived by expression (26)

\[\begin{gathered}
  {R_{k,ZF}} = \mathbb{E}\left[ {{{\log }_2}\left( {1 + SN{R_{k,BS,ZF}}} \right)} \right] \hfill \\
  \;\;\;\;\;\;\;\;\;\; = \mathbb{E}\left[ {{{\log }_2}\left\{ {1 + \frac{{SN{R_{UT}}}}{{{{\left[ {{{\left( {{{\mathbf{G}}^H}{\mathbf{G}}} \right)}^{ - 1}}} \right]}_{kk}}}}} \right\}} \right] \hfill \\
\end{gathered}, \tag{26} \]

Based on the channel matrix in (2), the closed-form solution of the lower bound of the ergodic achievable rate for the $k{\text{-th}}$ UT is obtained in $\textbf{\emph{Proposition 2}}$.

$\textbf{\emph{Proposition 2}}$: For a single-cell multi-user massive MIMO communication system with BS irregular antenna arrays and the zero-forcing detector scheme, the closed-form expression of the lower bound of the uplink ergodic achievable rate for the $k\text{-th}$ UT is given by expression (27a)

\[\begin{gathered}
  {{\overset{\lower0.5em\hbox{$\smash{\scriptscriptstyle\smile}$}}{R} }_{k,ZF}} =  \hfill \\
  {\log _2}\left\{ {1 + {{SN{R_{UT}}{\beta _k}} \mathord{\left/
 {\vphantom {{SN{R_{UT}}{\beta _k}} {\left[ {\Upsilon \sum\limits_{i = 1}^K {\sum\limits_{j = 2}^K {\Gamma \left( {j - 1} \right)} } } \right.}}} \right.
 \kern-\nulldelimiterspace} {\left[ {\Upsilon \sum\limits_{i = 1}^K {\sum\limits_{j = 2}^K {\Gamma \left( {j - 1} \right)} } } \right.}}} \right. \times  \hfill \\
  D\left( {i,j} \right)\left( {\tau _{n + i}^{n + j - 2}} \right. - \sum\limits_{p = 1}^n {\sum\limits_{q = 1}^n {{{\left[ {{\mathbf{B}}_{M,K}^{ - 1}} \right]}_{q,p}}\tau _{n + i}^{q - 1}\left. {\tau _p^{n + j - 2}} \right)} }  \hfill \\
   + \Upsilon \sum\limits_{i = 1}^K {D\left( {i,j} \right)} \left( {\tau _{n + i}^{n - 1}\left( {\ln \left( {{\tau _{n + i}}} \right) - \gamma } \right) - } \right. \hfill \\
  \left. {\left. {\left. {\sum\limits_{p = 1}^n {\sum\limits_{q = 1}^n {{{\left[ {{\mathbf{B}}_{M,K}^{ - 1}} \right]}_{q,p}}\tau _{n + i}^{q - 1}\tau _p^{n - 1}\left( {\ln \left( {{\tau _p}} \right) - \gamma } \right)} } } \right)} \right]} \right\} \hfill \\
\end{gathered}, \tag{27a} \]

with

\[\Upsilon  = \frac{{\det \left( {{{\mathbf{B}}_{M,K}}} \right)}}{{K\mathop \prod \limits_{q < p}^M \left( {{\tau _p} - {\tau _q}} \right)\mathop \prod \limits_{p = 1}^{K - 1} p!}},\tag{27b}\]
\[\begin{gathered}
{{\mathbf{B}}_{M,K}} = \left[ {\begin{array}{*{20}{c}}
  1&{{\tau _1}}& \cdots  \\
   \vdots & \vdots & \ddots  \\
  1&{{\tau _{M - K}}}& \ldots
\end{array}\;\;\;\begin{array}{*{20}{c}}
  {\tau _1^{M - K - 1}} \\
   \vdots  \\
  {\tau _{M - K}^{M - K - 1}}
\end{array}} \right] \hfill \\
= \left[ {\begin{array}{*{20}{c}}
  1&{{\tau _1}}& \cdots  \\
   \vdots & \vdots & \ddots  \\
  1&{{\tau _n}}& \ldots
\end{array}\;\;\;\begin{array}{*{20}{c}}
  {\tau _1^{n - 1}} \\
   \vdots  \\
  {\tau _n^{n - 1}}
\end{array}} \right]
\end{gathered}, \tag{27c}\]
where $n = M - K$, $\gamma$ is the Euler-Mascheroni constant, $D\left( {i,j} \right)$ is the cofactor of the element ${\left[ {\mathbf{\Omega }} \right]_{i,j}}$ in the $K \times K$ matrix ${\mathbf{\Omega }}$, and the element ${\left[ {\mathbf{\Omega }} \right]_{i,j}}$ is expressed by expression (27d)

\[\begin{gathered}
  {\left[ {\mathbf{\Omega }} \right]_{i,j}} = \left( {j - 1} \right)!\left( {\tau _{n + i}^{n + j - 1} - } \right. \hfill \\
  \left. {\;\;\sum\limits_{p = 1}^n {\sum\limits_{q = 1}^n {{{\left[ {{\mathbf{B}}_{M,K}^{ - 1}} \right]}_{p,q}}\tau _{n + i}^{p - 1}\tau _q^{n + j - 1}} } } \right) \hfill \\
\end{gathered}. \tag{27d} \]

$\textbf{\emph{Proof}}$:  Substituting (2) into (26) and assuming that eigenvalues of the channel correlation matrix are known, the achievable rate for the $k\text{-th}$ UT is derived by expression (28).

\[\begin{gathered}
  {R_{k,ZF}} = {\mathbb{E}_{\mathbf{H}}}\left[ {{{\log }_2}\left\{ {1 + \frac{{SN{R_{UT}}}}{{{{\left[ {{{\left( {{{\mathbf{G}}^H}{\mathbf{G}}} \right)}^{ - 1}}} \right]}_{kk}}}}} \right\}} \right] \hfill \\
   \geqslant {\log _2}\left\{ {1 + \frac{{SN{R_{UT}}}}{{{\mathbb{E}_{\mathbf{H}}}{{\left[ {{{\left( {{{\mathbf{G}}^H}{\mathbf{G}}} \right)}^{ - 1}}} \right]}_{kk}}}}} \right\} \hfill \\
   = {\log _2}\left[ {1 + \frac{{SN{R_{UT}}K{\beta _k}}}{{{\mathbb{E}_{\mathbf{H}}}\left\{ {{\text{trace}}\left[ {{{\left( {{{\mathbf{H}}^H}{{\mathbf{A}}^H}{{\mathbf{C}}^H}{\mathbf{CAH}}} \right)}^{ - 1}}} \right]} \right\}}}} \right] \hfill \\
   = {\log _2}\left\{ {1 + {{SN{R_{UT}}K{\beta _k}} \mathord{\left/
 {\vphantom {{SN{R_{UT}}K{\beta _k}} {\mathbb{E}\left( {\sum\limits_{i = 1}^K {\xi _i^{ - 1}} } \right)}}} \right.
 \kern-\nulldelimiterspace} {\mathbb{E}\left( {\sum\limits_{i = 1}^K {\xi _i^{ - 1}} } \right)}}} \right\} \hfill \\
\end{gathered}, \tag{28} \]
where ${\xi _i}$, $1 \leqslant i \leqslant K$, is the $i\text{-th}$ ordered eigenvalue of matrix ${{\mathbf{H}}^H}{{\mathbf{A}}^H}{{\mathbf{C}}^H}{\mathbf{CAH}}$. Let ${f_\xi }\left( {x|{\tau _1},...,{\tau _M}} \right)$ denote the conditional marginal PDF of the unordered eigenvalues of matrix ${{\mathbf{H}}^H}{{\mathbf{A}}^H}{{\mathbf{C}}^H}{\mathbf{CAH}}$, and based on results in \cite{Alfano04}, ${f_\xi }\left( {x|{\tau _1},...,{\tau _M}} \right)$ is expressed by expression (29)

\[\begin{gathered}
  {f_\xi }\left( {x|{\tau _1},...,{\tau _M}} \right) = \Upsilon \sum\limits_{i = 1}^K {\sum\limits_{j = 1}^K {{x^{j - 1}}D\left( {i,j} \right)} }  \times  \hfill \\
  \;\;\;\;\;\;\;\;\;\;\left( \begin{gathered}
  \tau _{n + i}^{n - 1}{e^{ - x/{\tau _{n + i}}}} -  \hfill \\
  \sum\limits_{p = 1}^n {\sum\limits_{q = 1}^n {{{\left[ {{\mathbf{B}}_{M,K}^{ - 1}} \right]}_{q,p}}\tau _{n + i}^{q - 1}\tau _p^{n - 1}{e^{ - x/{\tau _p}}}} }  \hfill \\
\end{gathered}  \right) \hfill \\
\end{gathered}. \tag{29} \]

Therefore, the term $\mathbb{E}\left( {\sum\limits_{i = 1}^K {\xi _i^{ - 1}} } \right)$ in (28) is derived by expression (30)

\[\begin{gathered}
  \mathbb{E}\left( {\sum\limits_{i = 1}^K {\xi _i^{ - 1}} } \right) = K\int\limits_{ - \infty }^{ + \infty } {\frac{1}{x}} {f_\xi }\left( {x|{\tau _1},...,{\tau _M}} \right){\text{d}}x \hfill \\
   = \Upsilon K\left\{ {\sum\limits_{i = 1}^K {\sum\limits_{j = 2}^K {\Gamma \left( {j - 1} \right)D\left( {i,j} \right) \times } } } \right. \hfill \\
  \left[ {\tau _{n + i}^{n + j - 2} - \sum\limits_{p = 1}^n {\sum\limits_{q = 1}^n {{{\left[ {{\mathbf{B}}_{M,K}^{ - 1}} \right]}_{q,p}}\tau _{n + i}^{q - 1}\tau _p^{n + j - 2}} } } \right] +  \hfill \\
  \sum\limits_{i = 1}^K {D\left( {i,j} \right)\left[ {\tau _{n + i}^{n - 1}\left( {\ln \left( {{\tau _{n + i}}} \right) - \gamma } \right) - } \right.}  \hfill \\
  \left. {\left. {\sum\limits_{p = 1}^n {\sum\limits_{q = 1}^n {{{\left[ {{\mathbf{B}}_{M,K}^{ - 1}} \right]}_{q,p}}\tau _{n + i}^{q - 1}\tau _p^{n - 1}\left( {\ln \left( {{\tau _p}} \right) - \gamma } \right)} } } \right]} \right\} \hfill \\
\end{gathered}. \tag{30} \]

\begin{figure*}[!t]

\[SE{R_{ZF}} = \frac{1}{K}\sum\limits_{k = 1}^K \begin{gathered}
  \frac{{{\omega _k}\Upsilon }}{{2\sqrt \pi  }}\sum\limits_{i = 1}^K {\sum\limits_{j = 1}^K {D\left( {i,j} \right)\left[ {\tau _{n + i}^{n - j - 1}G_{2,2}^{2,1}\left( {{\varpi _k}SN{R_{UT}}{\beta _k}{\tau _{n + i}}\left| \begin{gathered}
  1 - j,1 \hfill \\
  0,{1 \mathord{\left/
 {\vphantom {1 2}} \right.
 \kern-\nulldelimiterspace} 2} \hfill \\
\end{gathered}  \right.} \right)} \right.} }  \hfill \\
   - \sum\limits_{p = 1}^n {\sum\limits_{q = 1}^n {{{\left[ {{\mathbf{B}}_{M,K}^{ - 1}} \right]}_{q,p}}\tau _{n + i}^{q - 1}\tau _p^{n - j - 1}\left. {G_{2,2}^{2,1}\left( {{\varpi _k}SN{R_{UT}}{\beta _k}{\tau _{n + i}}\left| \begin{gathered}
  1 - j,1 \hfill \\
  0,{1 \mathord{\left/
 {\vphantom {1 2}} \right.
 \kern-\nulldelimiterspace} 2} \hfill \\
\end{gathered}  \right.} \right)} \right]} }  \hfill \\
\end{gathered}.\tag{33}  \]
\end{figure*}

Substituting (30) into (28), (28) is derived as

\[\begin{gathered}
  {R_{k,ZF}} \geqslant  \hfill \\
  {\log _2}\left\{ {1 + {{SN{R_{UT}}{\beta _k}} \mathord{\left/
 {\vphantom {{SN{R_{UT}}{\beta _k}} {\left[ {\Upsilon \sum\limits_{i = 1}^K {\sum\limits_{j = 2}^K {\Gamma \left( {j - 1} \right)} } } \right.}}} \right.
 \kern-\nulldelimiterspace} {\left[ {\Upsilon \sum\limits_{i = 1}^K {\sum\limits_{j = 2}^K {\Gamma \left( {j - 1} \right)} } } \right.}}} \right. \times  \hfill \\
  D\left( {i,j} \right)\left( {\tau _{n + i}^{n + j - 2}} \right. - \sum\limits_{p = 1}^n {\sum\limits_{q = 1}^n {{{\left[ {{\mathbf{B}}_{M,K}^{ - 1}} \right]}_{q,p}}\tau _{n + i}^{q - 1}\left. {\tau _p^{n + j - 2}} \right)} }  \hfill \\
   + \Upsilon \sum\limits_{i = 1}^K {D\left( {i,j} \right)} \left( {\tau _{n + i}^{n - 1}\left( {\ln \left( {{\tau _{n + i}}} \right) - \gamma } \right) - } \right. \hfill \\
  \left. {\left. {\left. {\sum\limits_{p = 1}^n {\sum\limits_{q = 1}^n {{{\left[ {{\mathbf{B}}_{M,K}^{ - 1}} \right]}_{q,p}}\tau _{n + i}^{q - 1}\tau _p^{n - 1}\left( {\ln \left( {{\tau _p}} \right) - \gamma } \right)} } } \right)} \right]} \right\} \hfill \\
\end{gathered}, \tag{31} \]
and the lower bound of the ergodic achievable rate of the $k$th UT is just at the right side of the sign of inequality.

Hence, $\textbf{\emph{Proposition 2}}$ gets proved. When all UTs are considered in the cell, the lower bound of the uplink ergodic achievable sum rate is derived by ${\overset{\lower0.5em\hbox{$\smash{\scriptscriptstyle\smile}$}}{R} _{BS}} = \sum\limits_{k = 1}^K {{{\overset{\lower0.5em\hbox{$\smash{\scriptscriptstyle\smile}$}}{R} }_{k,ZF}}} $. {From $\textbf{\emph{Proposition 2}}$, it can be intuitively found that the transmit SNR at the UT, the large scale fading coefficient and the number of UTs directly affect the lower bound of the ergodic achievable rate for the $k$th UT. With the higher transmit SNR at the UT or the higher path loss, the lower bound of the ergodic achievable rate is logarithmically increased or decreased, respectively. Moreover, the increasing of the number of UTs deteriorates the lower bound of the ergodic achievable rate. Because the eigenvalues of the matrix ${{\mathbf{H}}^H}{{\mathbf{A}}^H}{{\mathbf{C}}^H}{\mathbf{CAH}}$ are involved within (27) in a complicated form, it's hard to intuitively estimate the impact from the small scale fading coefficient, the mutual coupling effects and the array steering matrix. In the following numerical results will be provided to illustrate the variation trend of the lower bound of the ergodic achievable rate with the changing of these parameters.}

\subsection{Symbol error rate}
Considering the zero-forcing detector adopted at the BS, the average received SER of multi-user massive MIMO communication systems with irregular antenna arrays is expressed \cite{Proakis01}
\[SE{R_{ZF}} = \frac{1}{K}\sum\limits_{k = 1}^K {\mathbb{E}\left[ {{\omega _k}Q\left( {\sqrt {2{\varpi _k}SN{R_{k,BS,ZF}}} } \right)} \right]},\tag{32} \]
where $Q\left(  \cdot  \right)$ is the Gaussian $Q$ function while ${\omega _k}$ and ${\varpi _k}$ are modulation-specific constants. For the quadrature phase shift keying (QPSK) modulation, modulation-specific constants are configured as ${\omega _k} = 2$ and ${\varpi _k} = 0.5$. Assuming that all eigenvalues of channel correlation matrix ${\mathbf{\Psi }} = {\mathbf{CA}}{{\mathbf{A}}^H}{{\mathbf{C}}^H}$ are known, $\textbf{\emph{Proposition 3}}$ is obtained.

$\textbf{\emph{Proposition 3}}$: For a single-cell multi-user massive MIMO communication system with BS irregular antenna arrays and the zero-forcing detector scheme, the average received SER of multi-user massive MIMO communication systems is given by as expression (33), which locates at the beginning of this page.

$\textbf{\emph{Proof}}$: Substitute (25) into (32), the average received SER of multi-user massive MIMO communication systems with irregular antenna arrays is derived by expression (34)

\[\begin{gathered}
  SE{R_{ZF}} = \frac{1}{K}\sum\limits_{k = 1}^K {\frac{{{\omega _k}}}{2} \times }  \hfill \\
  {\mathbb{E}_{\mathbf{H}}}\left[ {erfc\left( {\sqrt {\frac{{{\varpi _k}SN{R_{UT}}}}{{{{\left[ {{{\left( {{{\mathbf{G}}^H}{\mathbf{G}}} \right)}^{ - 1}}} \right]}_{kk}}}}} } \right)} \right] \hfill \\
   = \frac{1}{K}\sum\limits_{k = 1}^K {\frac{{{\omega _k}}}{2}{\mathbb{E}_{\mathbf{H}}}\left[ {erfc\left( {\sqrt {{\varpi _k}SN{R_{UT}}{\beta _k}\xi } } \right)} \right]}  \hfill \\
\end{gathered}, \tag{34} \]

where $erfc\left(  \cdot  \right)$ is the complementary error function which can be expressed by Meijer's G-function as $erfc\left( {\sqrt x } \right) = \frac{1}{\pi }G_{1,2}^{2,0}\left( {x\left| {\begin{array}{*{20}{l}}
  1 \\
  {0,{1 \mathord{\left/
 {\vphantom {1 2}} \right.
 \kern-\nulldelimiterspace} 2}}
\end{array}} \right.} \right)$ \cite[Eq. (8.4.14.2)]{Humar11}. Furthermore, the average received SER of multi-user massive MIMO communication systems with irregular antenna arrays is rewritten by expression (35)

\[\begin{gathered}
  SE{R_{ZF}} = \frac{1}{K}\sum\limits_{k = 1}^K {\left[ {\frac{{{\omega _k}}}{2}\int_0^\infty  {\left( {\frac{1}{\pi } \times } \right.} } \right.}  \hfill \\
  G_{1,2}^{2,0}\left( {{\varpi _k}SN{R_{UT}}{\beta _k}x\left| \begin{gathered}
  1 \hfill \\
  0,{1 \mathord{\left/
 {\vphantom {1 2}} \right.
 \kern-\nulldelimiterspace} 2} \hfill \\
\end{gathered}  \right.} \right) \times  \hfill \\
  \left. {\left. {{f_\xi }\left( {x|{\tau _1},...,{\tau _M}} \right)dx} \right)} \right] \hfill \\
\end{gathered}. \tag{35} \]

Substitute (29) into (35), the average received SER of multi-user massive MIMO communication systems with irregular antenna arrays in (33) is completed by replacing the integral expression in \cite[Eq. (7.831)]{Jeffrey07}. {From (33) it can be intuitively found that the average received SER decreases with the increasing of the number of UTs even after equalization. But similar to (27), the eigenvalues of the matrix ${{\bf{H}}^H}{{\bf{A}}^H}{{\bf{C}}^H}{\bf{CAH}}$ are still complexly involved within the expression of the average received SER. Therefore, the impact of the small scale fading, mutual coupling and spatial correlation on the average received SER is difficult to be intuitively estimated. In the following it will be shown that the impact of the small scale fading will be neglected after assuming the number of antennas growing without bound. Numerical results will illustrate how the mutual coupling and spatial correlation influence the average received SER.}

\subsection{Outage probability}
The outage probability is one of the most important metrics for wireless communication systems. Assuming the SNR threshold is given by $SN{R_{{\text{th}}}}$, the average outage probability of multi-user massive MIMO communication systems with irregular antenna arrays is defined as \cite{Richter09}
\[{P_{{\text{out}}}} = \frac{1}{K}\sum\limits_{k = 1}^K {\Pr \left( {SN{R_{k,BS,ZF}} \leqslant SN{R_{{\text{th}}}}} \right)}.\tag{36} \]

Based on the scenario illustrated in Fig. 1, $\textbf{\emph{Proposition 4}}$ is obtained as following.

$\textbf{\emph{Proposition 4}}$: For a single-cell multi-user massive MIMO communication system with BS irregular antenna arrays and the zero-forcing detector scheme, the average outage probability of multi-user massive MIMO communication systems with is given by expression (37)

\[\begin{gathered}
  {P_{{\text{out}}}} = \frac{1}{K}\sum\limits_{k = 1}^K {\Upsilon \sum\limits_{i = 1}^K {\sum\limits_{j = 1}^K {D\left( {i,j} \right) \times } } }  \hfill \\
  \left( \begin{gathered}
  \vartheta \left( {{\tau _{n + i}},j,k} \right) \hfill \\
   - \sum\limits_{p = 1}^n {\sum\limits_{q = 1}^n {{{\left[ {{\mathbf{B}}_{M,K}^{ - 1}} \right]}_{q,p}}\tau _{n + i}^{q - 1}\vartheta \left( {{\tau _p},j} \right)} }  \hfill \\
\end{gathered}  \right) \hfill \\
\end{gathered}, \tag{37a} \]

with

\[\begin{gathered}
  \vartheta \left( {x,y,k} \right) = \left( {y - 1} \right)!{x^{n + y - 1}} -  \hfill \\
  \;\;\;\;\;\;\;\exp \left( { - \frac{{SN{R_{th}}}}{{SN{R_{UT}}{\beta _k}x}}} \right) \times  \hfill \\
  \;\;\;\;\;\;\mathop \sum \limits_{s = 0}^{y - 1} \frac{{\left( {y - 1} \right)!}}{{s!}}{\left( {\frac{{SN{R_{th}}}}{{SN{R_{UT}}{\beta _k}}}} \right)^s}{x^{n + y - s - 3}} \hfill \\
\end{gathered}, \tag{37b} \]

$\textbf{\emph{Proof}}$: Substitute (25) into (36), the average outage probability of multi-user massive MIMO communication systems is derived by
\[\begin{gathered}
  {P_{{\text{out}}}} = \frac{1}{K}\sum\limits_{k = 1}^K {\Pr \left( {\frac{{SN{R_{UT}}}}{{{{\left[ {{{\left( {{{\mathbf{G}}^H}{\mathbf{G}}} \right)}^{ - 1}}} \right]}_{kk}}}} \leqslant SN{R_{{\text{th}}}}} \right)}  \hfill \\
  \;\;\;\;\;\; = \frac{1}{K}\sum\limits_{k = 1}^K {\Pr \left( {SN{R_{UT}}{\beta _k}\xi  \leqslant SN{R_{{\text{th}}}}} \right)}  \hfill \\
  \;\;\;\;\;\; = \frac{1}{K}\sum\limits_{k = 1}^K {\Pr \left( {\xi  \leqslant \frac{{SN{R_{{\text{th}}}}}}{{SN{R_{UT}}{\beta _k}}}} \right)}  \hfill \\
\end{gathered}.\tag{38} \]

Assuming that all eigenvalues of channel correlation matrix ${\mathbf{\Psi }} = {\mathbf{CA}}{{\mathbf{A}}^H}{{\mathbf{C}}^H}$ are known, (38) is rewritten by

\[{P_{{\text{out}}}} = \frac{1}{K}\sum\limits_{k = 1}^K {\int_0^{\frac{{SN{R_{{\text{th}}}}}}{{SN{R_{UT}}{\beta _k}}}} {{f_\xi }\left( {x|{\tau _1},...,{\tau _M}} \right)dx} }.\tag{39} \]

Substitute (29) into (38), the average outage probability of multi-user massive MIMO communication systems in (36) is completed by replacing the integral expression in \cite[Eq. (3.351.8)]{Jeffrey07}. {For the estimation of the system outage performance, the setting of the SNR threshold is very critical in practical systems. Equation (37) implies that the average outage probability monotonously increases with the increas of the SNR threshold. When the number of UTs is increased, the average outage probability also increases. Numerical results will be provided in the next section to show how the number of BS antennas and antenna array size affect the average outage probability.}

\subsection{Asymptotic analysis}

To further investigate asymptotic results on multi-user massive MIMO communication systems with irregular antenna arrays, we then assume the scenario where the number of antennas $M$ grows without bound, i.e. $M \to \infty $. In such a scenario, the following expressions will hold \cite{Hoydis13}

\[{{\mathbf{p}}^H}{\mathbf{Rp}} - \frac{1}{M}{\text{trace}}\left( {\mathbf{R}} \right)\xrightarrow[{M \to \infty }]{}0, \tag{40a}\]

\[{{\mathbf{p}}^H}{\mathbf{Rq}}\xrightarrow[{M \to \infty }]{}0, \tag{40b}\]
where matrix ${\mathbf{R}} \in {\mathbb{C}^{M \times M}}$ has uniformly bounded spectral norm with respect to $M$. And ${\mathbf{p}},{\mathbf{q}} \in {\mathbb{C}^{M \times 1}}$ are two independent vectors with distributions ${\mathbf{p}},{\mathbf{q}} \sim \mathcal{C}\mathcal{N}{\text{(0,}}\frac{1}{M}{{\mathbf{I}}_M}{\text{)}}$. $\mathbf{q}$ and $\mathbf{p}$ are also independent from $\mathbf{R}$. With the above expressions, we have the following proposition for the received SNR at the BS when the zero-forcing detector is adopted.

$\textbf{\emph{Proposition 5}}$: For a single cell multi-user massive MIMO communication system with BS irregular antenna arrays and the zero-forcing detector, when the number of antennas at the BS approaches infinity, i.e., $M \to \infty $, the received SNR over the link of the $k$th UT is expressed as
\[SN{R_{k,BS,ZF}}\xrightarrow[{M \to \infty }]{}SN{R_{UT}}{\beta _k}{\text{trace}}\left( {{{\mathbf{A}}^H}{{\mathbf{C}}^H}{\mathbf{CA}}} \right). \tag{41}\]

$\textbf{\emph{Proof}}$: The BS received SNR over the link of the $k$th UT is expressed by (25). Multiply the numerator and denominator by $M$, we get
\[SN{R_{k,BS,ZF}} = \frac{{M \times SN{R_{UT}}}}{{{{\left[ {{{\left( {\frac{1}{M}{{\mathbf{G}}^H}{\mathbf{G}}} \right)}^{ - 1}}} \right]}_{kk}}}}. \tag{42}\]

For the channel matrix multiplication term $\frac{1}{M}{{\mathbf{G}}^H}{\mathbf{G}}$, we have expression (43)

\[\begin{gathered}
  \frac{1}{M}{{\mathbf{G}}^H}{\mathbf{G}} = \frac{1}{M}{{\mathbf{D}}^H}{{\mathbf{H}}^H}{{\mathbf{A}}^H}{{\mathbf{C}}^H}{\mathbf{CAHD}} \hfill \\
  {\kern 1pt}  = \frac{1}{M}\left[ {\begin{array}{*{20}{c}}
  \begin{gathered}
  \beta _1^{{1 \mathord{\left/
 {\vphantom {1 2}} \right.
 \kern-\nulldelimiterspace} 2}}{\mathbf{h}}_1^H \hfill \\
   \vdots  \hfill \\
  \beta _1^{{1 \mathord{\left/
 {\vphantom {1 2}} \right.
 \kern-\nulldelimiterspace} 2}}{\mathbf{h}}_k^H \hfill \\
   \vdots  \hfill \\
\end{gathered}  \\
  {\beta _1^{{1 \mathord{\left/
 {\vphantom {1 2}} \right.
 \kern-\nulldelimiterspace} 2}}{\mathbf{h}}_K^H}
\end{array}} \right]{{\mathbf{A}}^H}{{\mathbf{C}}^H}{\mathbf{CA}}{\left[ {\begin{array}{*{20}{c}}
  \begin{gathered}
  \beta _1^{{1 \mathord{\left/
 {\vphantom {1 2}} \right.
 \kern-\nulldelimiterspace} 2}}{{\mathbf{h}}_1} \hfill \\
   \vdots  \hfill \\
  \beta _1^{{1 \mathord{\left/
 {\vphantom {1 2}} \right.
 \kern-\nulldelimiterspace} 2}}{{\mathbf{h}}_k} \hfill \\
   \vdots  \hfill \\
\end{gathered}  \\
  {\beta _1^{{1 \mathord{\left/
 {\vphantom {1 2}} \right.
 \kern-\nulldelimiterspace} 2}}{{\mathbf{h}}_K}}
\end{array}} \right]^T} \hfill \\
\end{gathered}. \tag{43} \]

\begin{figure*}[!t]

\[\begin{gathered}
  SE{R_{ZF}}\xrightarrow[{M \to \infty }]{} = \frac{1}{K}\sum\limits_{k = 1}^K {\frac{{{\omega _k}}}{2}erfc\left( {\sqrt {{\varpi _k}SN{R_{UT}}{\beta _k}{\text{trace}}\left( {{{\mathbf{A}}^H}{{\mathbf{C}}^H}{\mathbf{CA}}} \right)} } \right)}  \hfill \\
  \;\;\;\;\;\;\;\; = \frac{1}{{2\pi K}}\sum\limits_{k = 1}^K {{\omega _k}G_{1,2}^{2,0}\left( {{\varpi _k}SN{R_{UT}}{\beta _k}{\text{trace}}\left( {{{\mathbf{A}}^H}{{\mathbf{C}}^H}{\mathbf{CA}}} \right)\left| {\begin{array}{*{20}{l}}
  1 \\
  {0,{1 \mathord{\left/
 {\vphantom {1 2}} \right.
 \kern-\nulldelimiterspace} 2}}
\end{array}} \right.} \right)}  \hfill \\
\end{gathered}. \tag{46}\]

\[SN{R_{k,BS,MRC}} = \frac{{SN{R_{UT}}{{\left| {{\beta _k}{\mathbf{h}}_k^H{{\mathbf{A}}^H}{{\mathbf{C}}^H}{\mathbf{CA}}{{\mathbf{h}}_k}} \right|}^2}}}{{SN{R_{UT}}\sum\limits_{i = 1,i \ne k}^K {{{\left| {\beta _k^{{1 \mathord{\left/
 {\vphantom {1 2}} \right.
 \kern-\nulldelimiterspace} 2}}{\mathbf{h}}_k^H{{\mathbf{A}}^H}{{\mathbf{C}}^H}{\mathbf{CA}}{{\mathbf{h}}_i}\beta _i^{{1 \mathord{\left/
 {\vphantom {1 2}} \right.
 \kern-\nulldelimiterspace} 2}}} \right|}^2} + {\beta _k}{\mathbf{h}}_k^H{{\mathbf{A}}^H}{{\mathbf{C}}^H}{\mathbf{CA}}{{\mathbf{h}}_k}} }}. \tag{47}\]

\end{figure*}

It can be seen that the entry at the $i$th row and $j$th column of $\frac{1}{M}{{\mathbf{G}}^H}{\mathbf{G}}$ is expressed by $\frac{{\sqrt {{\beta _i}{\beta _j}} }}{M}{\mathbf{h}}_i^H{{\mathbf{A}}^H}{{\mathbf{C}}^H}{\mathbf{CA}}{{\mathbf{h}}_j}$. If $i \ne j$, applying (40), we have $\frac{{\sqrt {{\beta _i}{\beta _j}} }}{M}{\mathbf{h}}_i^H{{\mathbf{A}}^H}{{\mathbf{C}}^H}{\mathbf{CA}}{{\mathbf{h}}_j}\xrightarrow[{M \to \infty }]{}0$. If $i=j$, we have $\frac{{{\beta _i}}}{M}{\mathbf{h}}_i^H{{\mathbf{A}}^H}{{\mathbf{C}}^H}{\mathbf{CA}}{{\mathbf{h}}_i}\xrightarrow[{M \to \infty }]{}\frac{{{\beta _i}}}{M}{\text{trace}}\left( {{{\mathbf{A}}^H}{{\mathbf{C}}^H}{\mathbf{CA}}} \right)$. In other words, when the number of antennas at the BS approaches infinity, $\frac{1}{M}{{\mathbf{G}}^H}{\mathbf{G}}$ converges to a diagonal matrix whose $i$th diagonal entry is $\frac{{{\beta _i}}}{M}{\text{trace}}\left( {{{\mathbf{A}}^H}{{\mathbf{C}}^H}{\mathbf{CA}}} \right)$. Therefore, the received SNR over the link of the $k$th UT is derived by expression (44)

\[\begin{gathered}
  SN{R_{k,BS,ZF}} =  \hfill \\
  \frac{{M \times SN{R_{UT}}}}{{{{\left[ {{{\left( {\frac{1}{M}{{\mathbf{G}}^H}{\mathbf{G}}} \right)}^{ - 1}}} \right]}_{kk}}}}\xrightarrow[{M \to \infty }]{}\frac{{M \times SN{R_{UT}}}}{{\frac{M}{{{\beta _i}{\text{trace}}\left( {{{\mathbf{A}}^H}{{\mathbf{C}}^H}{\mathbf{CA}}} \right)}}}} \hfill \\
   = SN{R_{UT}}{\beta _k}{\text{trace}}\left( {{{\mathbf{A}}^H}{{\mathbf{C}}^H}{\mathbf{CA}}} \right) \hfill \\
\end{gathered}. \tag{44} \]

So $\textbf{\emph{Proposition 5}}$ is proved.

$\textbf{\emph{Proposition 5}}$ illustrates that when the number of antennas at the BS approaches infinity, the received SNR only depends on the UT transmit power, large scale fading and channel correlation, despite of the small scale fading and noise. This result is very concise and coincides with the precious studies on massive MIMO \cite{Marzetta10}, \cite{Ngo13}.

{With the derived asymptotic results of the received SNR, the asymptotic result for the achievable sum rate, which is previously investigated in $\textbf{\emph{Proposition 2}}$, is given by} expression (45)

\[\begin{gathered}
  {R_{BS}} = \sum\limits_{k = 1}^K {{R_{k,ZF}}} \xrightarrow[{M \to \infty }]{{}}\sum\limits_{k = 1}^K {{{\log }_2}} \left[ {1 + } \right. \hfill \\
  \left. { SN{R_{UT}}{\beta _k}{\text{trace}}\left( {{{\mathbf{A}}^H}{{\mathbf{C}}^H}{\mathbf{CA}}} \right)} \right] \hfill \\
\end{gathered}. \tag{45} \]

{For the average SER investigated in Proposition 3, i.e., substituting (41) into (32), its asymptotic result when the number of BS antennas grows without a bound is given by} expression (46).

It is worth nothing that if the MRC detector is employed at the BS, the received SNR over the link of the $k$th UT can be expressed as expression (47).

When the number of antennas grows without a bound, similar to the proof of $\textbf{\emph{Proposition 5}}$, the numerator of (47) converges to $SN{R_{UT}}{\left| {{\beta _k}{\text{trace}}\left( {{{\mathbf{A}}^H}{{\mathbf{C}}^H}{\mathbf{CA}}} \right)} \right|^2}$ and the denominator of (47) converges to ${\beta _k}{\text{trace}}\left( {{{\mathbf{A}}^H}{{\mathbf{C}}^H}{\mathbf{CA}}} \right)$. Therefore, the received SNR over the link of the $k$th UT goes to $SN{R_{UT}}{\beta _k}{\text{trace}}\left( {{{\mathbf{A}}^H}{{\mathbf{C}}^H}{\mathbf{CA}}} \right)$ when the number of antennas approaches infinity. It can be found that the MRC detector and ZF detector have identical asymptotical results when the number of antennas approaches infinity. {With the asymptotic result of the received SNR when the MRC detector is employed, the asymptotic result for the ergodic received gain, which is investigated in Proposition 1, is obtained as}
\[\begin{array}{l}
\left( {M,R} \right) = SN{R_{UT}}{\beta _k}{\rm{trace}}\left( {{{\bf{A}}^H}{{\bf{C}}^H}{\bf{CA}}} \right)\\
\;\;\;\;\;\;\;\;\;\;\; - SN{R_{UT}}{\beta _k}{\rm{trace}}\left( {{{{\bf{\hat A}}}^H}{{{\bf{\hat C}}}^H}{\bf{\hat C\hat A}}} \right)
\end{array}. \tag{48}\]

\section{Simulation Results and Discussions}
Based on the proposed models of massive MIMO communication systems, the effect of mutual coupling on the massive MIMO communication systems with irregular and regular antenna arrays is analyzed by numerical simulations. In the following, some default parameters and assumptions are specified. The type of all antennas at the BS is assumed to be the same, and the load impedance and self-impedance of each antenna are assumed to be 50 Ohms \cite{Masouros13}. The number of UTs in a cell is $K=10$ and the large scale fading factor ${\beta _k}$ is modeled as ${\beta _k} = {{z \mathord{\left/
 {\vphantom {z {\left( {{{{l_k}} \mathord{\left/
 {\vphantom {{{l_k}} {{l_{{\text{resist}}}}}}} \right.
 \kern-\nulldelimiterspace} {{l_{{\text{resist}}}}}}} \right)}}} \right.
 \kern-\nulldelimiterspace} {\left( {{{{l_k}} \mathord{\left/
 {\vphantom {{{l_k}} {{l_{{\text{resist}}}}}}} \right.
 \kern-\nulldelimiterspace} {{l_{{\text{resist}}}}}}} \right)}}^v}$, which is similar to ${\beta _1}$ except with ${l_k}$ being a uniformly distributed random variable ranging from 10 m to 150 m \cite{Ngo13J,Ge15}. The transmitting SNR at each UT is assumed to be 15 dB \cite{Shen10}. Since the BS is assumed to be associated with a large but finite number of independent incident directions, the number of incident directions $P$ are assumed to be 100 \cite{Masouros13,EnergyGe15}.

 {Fig. 5 illustrates the uplink ergodic achievable sum rate with respect to the number of antennas $M$, the circle radius $R$ and the irregularity coefficient $\zeta$ of antenna arrays. The lines correspond to the lower bound of the uplink ergodic achievable sum rate } ${\overset{\lower0.5em\hbox{$\smash{\scriptscriptstyle\smile}$}}{R} _{BS}}$. {The square points correspond to the asymptotic results of the achievable sum rate obtained in (45). When the circle radius is fixed, numerical results demonstrate that there exists a maximum of the uplink ergodic achievable sum rate with the increasing number of antennas. The uplink ergodic achievable sum rate increases with the increase of the number of antennas before achieving the maximum. After the number of antennas exceeds a given threshold, the uplink ergodic achievable sum rate becomes to decrease. When the number of antennas is fixed, the uplink ergodic achievable sum rate increases with the increase of the circle radius. Furthermore, it can be seen that the smaller value of the irregularity coefficient corresponds to the larger value of the achievable sum rate when the number of antennas is less than a given threshold. But when the number of antennas increases, curves with different irregularity coefficients get crossed. When the number of antennas is larger than a given threshold, the smaller value of the irregularity coefficient corresponds to the smaller value of the achievable sum rate. These results indicate that the irregular antenna array has contributed to improve the uplink ergodic achievable sum rate when the number of antennas is larger than or equal to a given threshold. In addition, the asymptotic results well match the lower bound of the uplink ergodic achievable sum rate in Fig. 5, especially when the number of antennas is large.}
\begin{figure}
\vspace{0.1in}
\centerline{\includegraphics[width=9.5cm,draft=false]{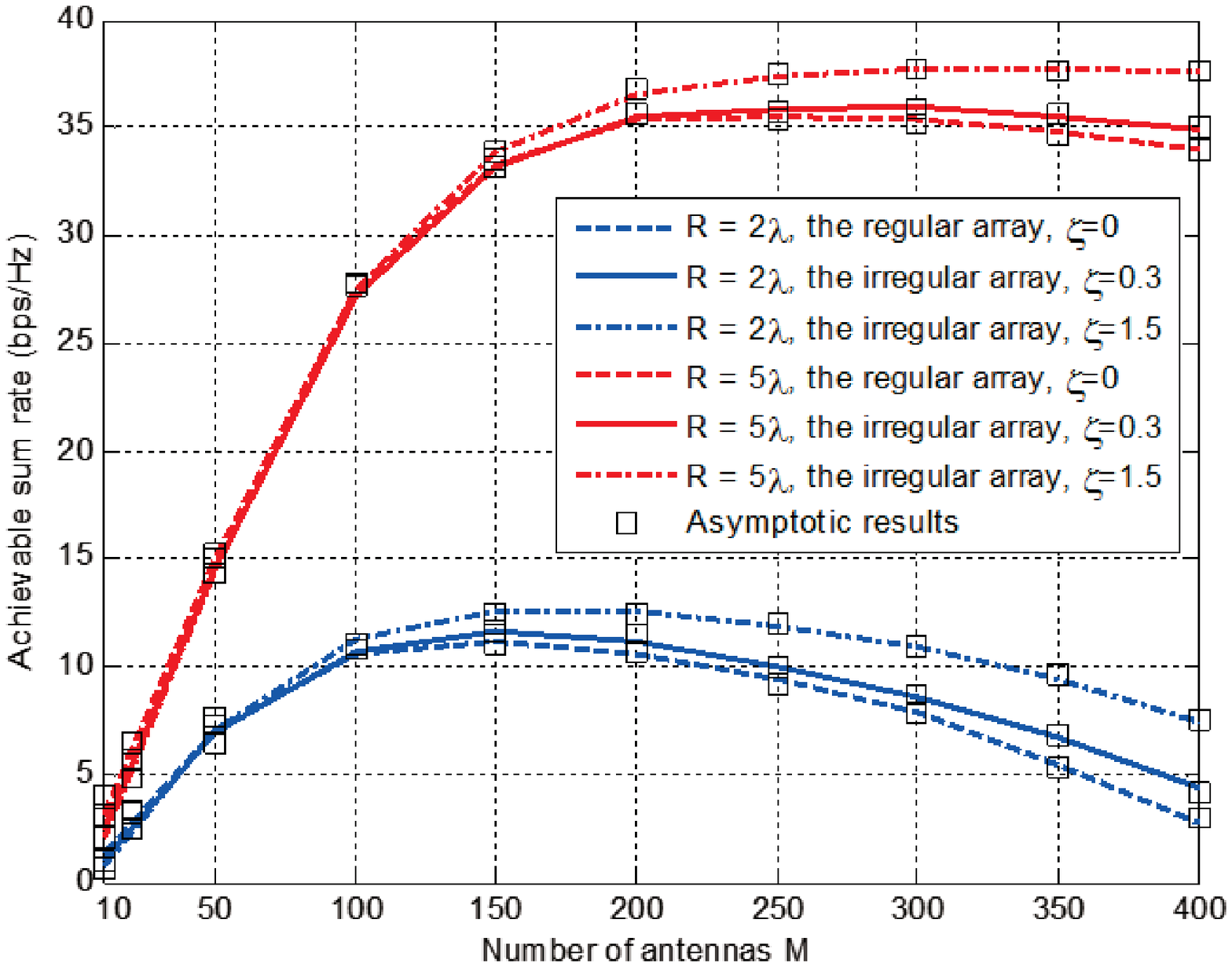}}
{\small Fig. 5. \@ \@ The achievable sum rate with respect to the number of antennas, the circle radius and the irregularity coefficient.}
\end{figure}

\begin{figure}
\vspace{0.1in}
\centerline{\includegraphics[width=9.5cm,draft=false]{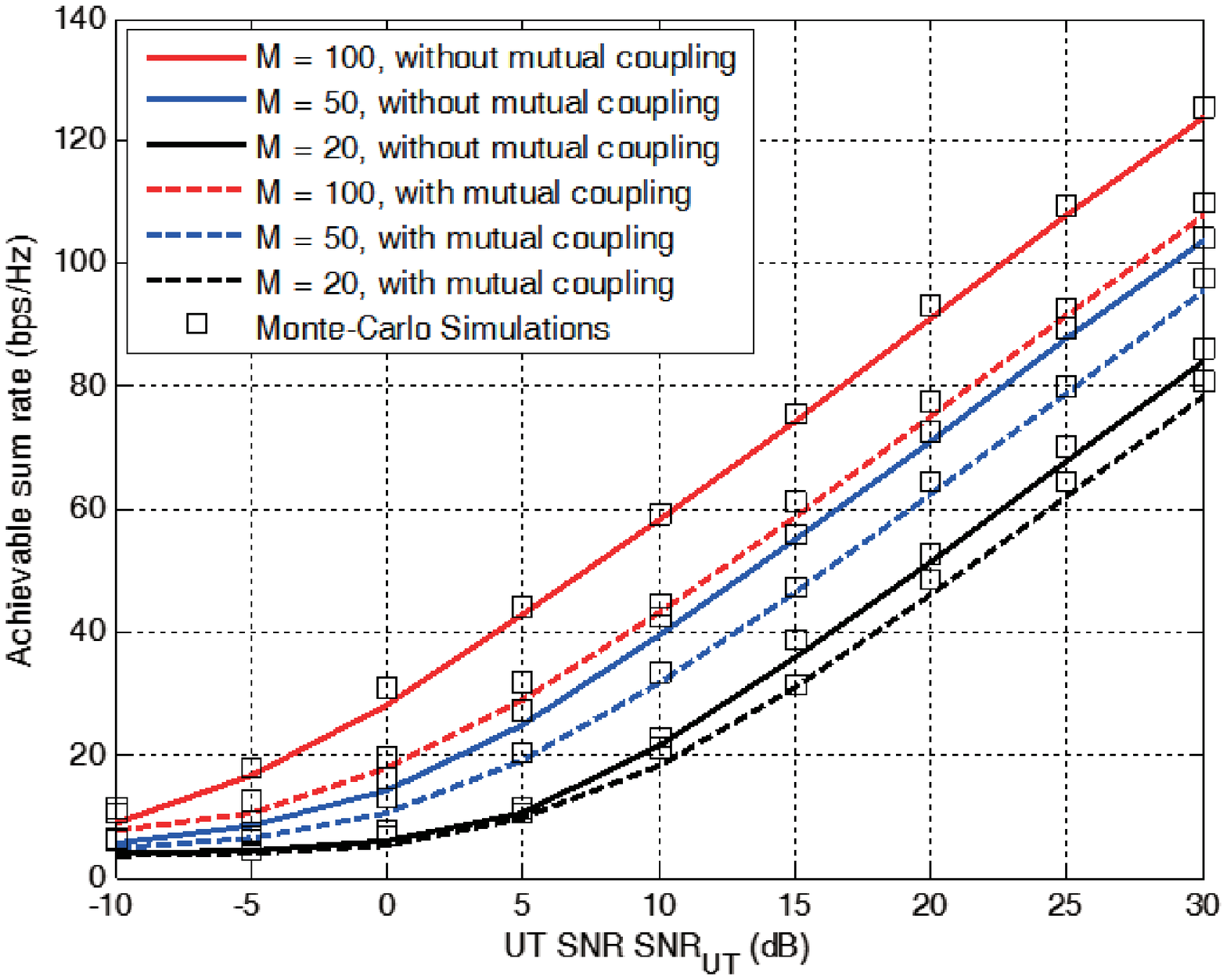}}
{\small Fig. 6. \@ \@ Impact of the UT SNR and the number of BS antennas on the achievable sum rate of multi-user massive MIMO communication systems with and without mutual coupling.}
\end{figure}
 Impact of the UT SNR and the number of BS antennas on the lower bound of the uplink ergodic achievable sum rate of multi-user massive MIMO communication systems with and without mutual coupling is investigated in Fig. 6. When the number of antennas at the BS is fixed, the lower bound of the uplink ergodic achievable sum rate increases with the increase of the UT SNR. When the UT SNR is fixed, the lower bound of the uplink ergodic achievable sum rate increases with the increase of the number of BS antennas. Moreover, the results of 10000 times Monte-Carlo simulation on the uplink ergodic achievable sum rate are illustrated. Both numerical and Monte-Carlo simulation results demonstrate that the uplink ergodic achievable sum rate with mutual coupling is less than the uplink ergodic achievable sum rate without mutual coupling for multi-user massive MIMO communication systems with irregular antenna arrays.

 Without loss of generality, the QPSK modulation scheme is adopted for numerical simulations in Fig. 7. The modulation constants are configured as ${\omega _k} = 2$ and ${\varpi _k} = 0.5$. Impact of the UT SNR, the number of BS antennas $M$ and the circle radius $R$ on the average SER of multi-user massive MIMO communication systems with irregular antenna arrays is evaluated in Fig. 7. The solid and dashed lines corresponds to the SER with limited number of antennas under different circle radii. And the square points correspond to the asymptotic results obtained in (46). When the number of BS antennas and the circle radius are fixed, the average SER decreases with the increase of the UT SNR. When the UT SNR and the circle radius are fixed, the average SER decreases with the increase of the number of BS antennas. When the UT SNR and the number of BS antennas are fixed, the average SER decreases with the increase of the circle radius.

\begin{figure}
\vspace{0.1in}
\centerline{\includegraphics[width=9.5cm,draft=false]{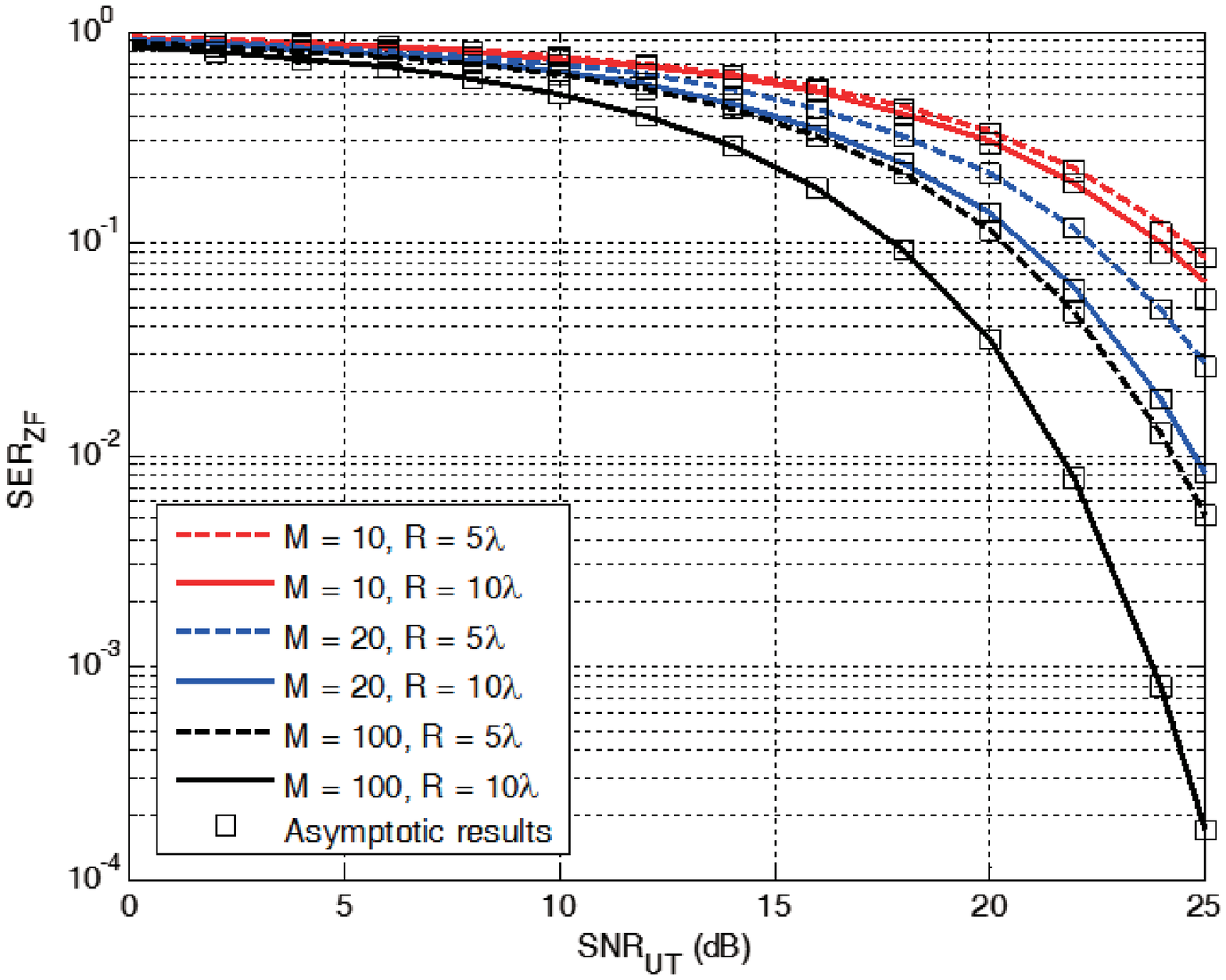}}
{\small Fig. 7. \@ \@ The average SER with respect to the UT SNR, the number of BS antennas and the circle radius of the massive MIMO communication system with irregular antenna arrays. Both cases with limited and infinite number of antennas are illustrated.}
\end{figure}
\begin{figure}
\vspace{0.1in}
\centerline{\includegraphics[width=9.5cm,draft=false]{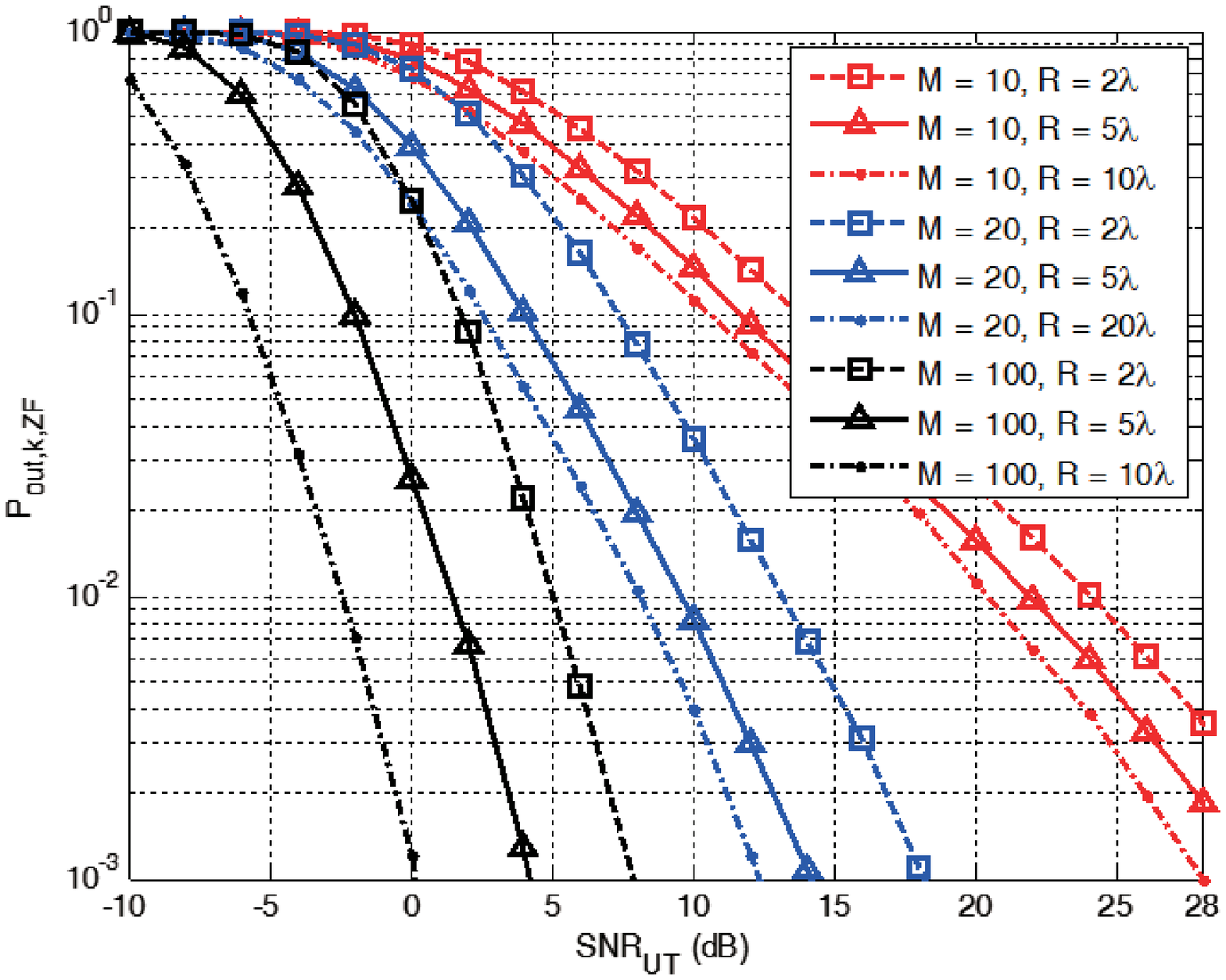}}
{\small Fig. 8. \@ \@ The average outage probability with respect to the UT SNR, the number of BS antennas and the circle radius of the massive MIMO communication system with irregular antenna arrays.}
\end{figure}

 The average outage probability with respect to the UT SNR, the number of BS antennas and the circle radius is analyzed in Fig. 8. Without loss of generality, the SNR threshold is configured as $SN{R_{{\text{th}}}} = -3$ dB. When the number of BS antennas and the circle radius are fixed, the average outage probability of massive MIMO communication systems with irregular antenna arrays decreases with the increase of the UT SNR. When the UT SNR and the circle radius are fixed, the average outage probability of massive MIMO communication systems with irregular antenna arrays decreases with the increase of the number of BS antennas. When the UT SNR and the number of BS antennas are fixed, the average outage probability of massive MIMO communication systems with irregular antenna arrays decreases with the increase of the circle radius.

\section{Conclusion}
In this paper, multi-user massive MIMO communication systems with irregular antenna arrays and mutual coupling effects have been investigated. In real antenna deployment scenarios, antenna spatial distances of massive MIMO antenna arrays are usually irregular. Considering engineering requirements from real scenarios, the effect of the mutual coupling on the irregular antenna array is firstly analyzed by the channel correlation model and ergodic received gain. Furthermore, the lower bound of the ergodic achievable rate, the average SER and the average outage probability of multi-user massive MIMO communication systems with irregular antenna arrays are proposed. Numerical results indicate that there exists a maximum for the achievable rate considering different numbers of antennas for massive MIMO communication systems. Compared with the regular antenna array, the irregular antenna array has contributed to improve the achievable rate when the number of antennas is larger than or equal to a specific threshold. Our results provide some guidelines for the massive MIMO antenna deployment in real scenarios. For the future study, we will try to investigate multi-cell multi-user massive MIMO communication systems with irregular antenna arrays.

%\newpage

\begin{IEEEbiography}[{\includegraphics[width=1in,height=1.25in,clip,keepaspectratio]{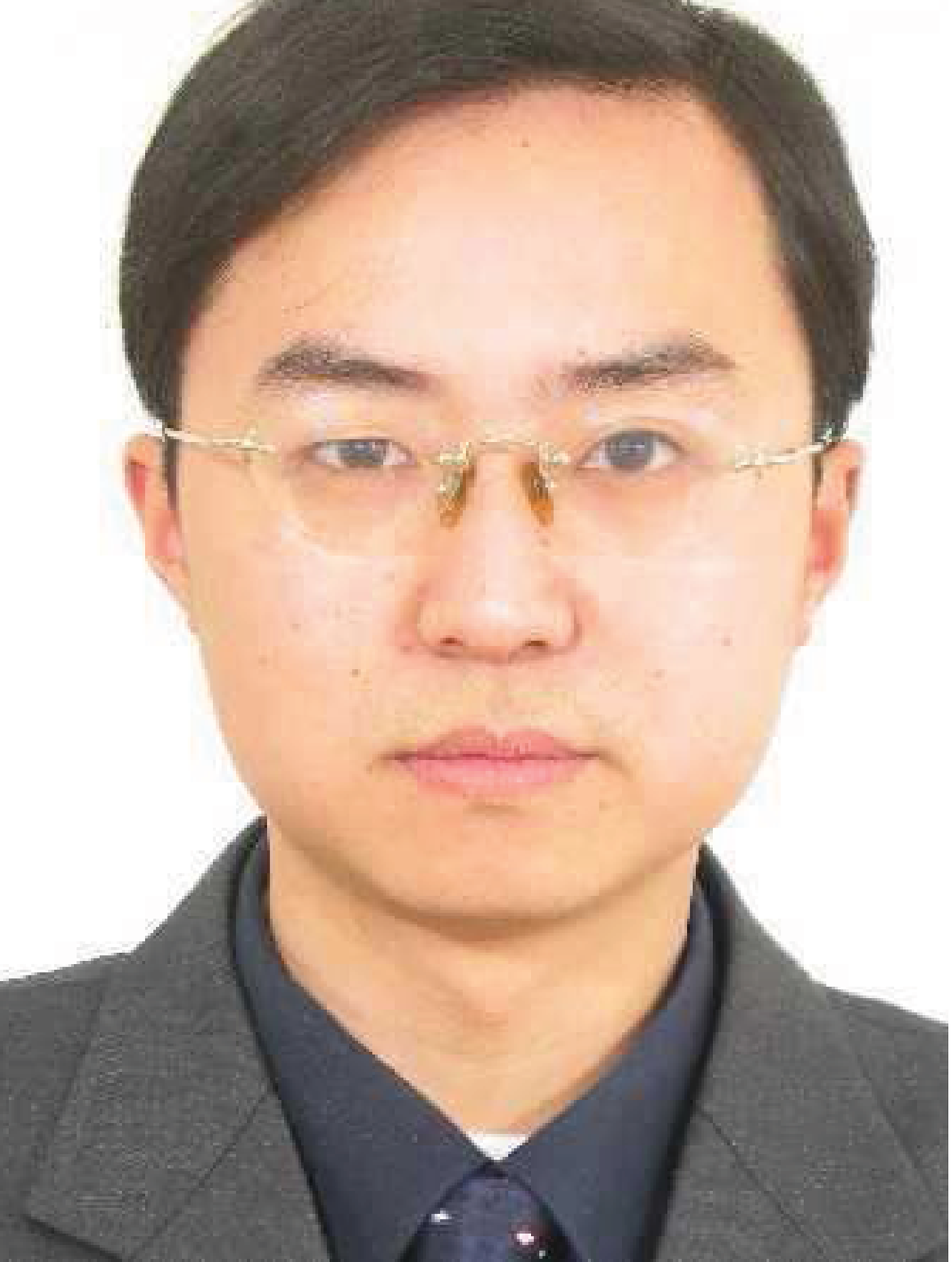}}]{Xiaohu~Ge}
(M'09-SM'11) is currently a full Professor with the School of Electronic Information and Communications at Huazhong University of Science and Technology (HUST), China. He is an adjunct professor with with the Faculty of Engineering and Information
Technology at University of Technology Sydney (UTS), Australia. He received his PhD degree in Communication and Information Engineering from HUST in 2003. He has worked at HUST since Nov. 2005. Prior to that, he worked as a researcher at Ajou University (Korea) and Politecnico Di Torino (Italy) from Jan. 2004 to Oct. 2005. His research interests are in the area of mobile communications, traffic modeling in wireless networks, green communications, and interference modeling in wireless communications. He has published more than 100 papers in refereed journals and conference proceedings and has been granted about 15 patents in China. He received the Best Paper Awards from IEEE Globecom 2010. 

Dr. Ge is a Senior Member of the China Institute of Communications and a member of the National Natural Science Foundation of China and the Chinese Ministry of Science and Technology Peer Review College. He has been actively involved in organizing more the ten international conferences since 2005. He served as the general Chair for the 2015 IEEE International Conference on Green Computing and Communications (IEEE GreenCom 2015). He serves as an Associate Editor for the \textit{IEEE ACCESS}, \textit{Wireless Communications and Mobile Computing Journal (Wiley)} and \textit{the International Journal of Communication Systems (Wiley)}, etc. Moreover, he served as the guest editor for \textit{IEEE Communications Magazine} Special Issue on 5G Wireless Communication Systems.
\end{IEEEbiography}

\begin{IEEEbiography}[{\includegraphics[width=1in,height=1.25in,clip,keepaspectratio]{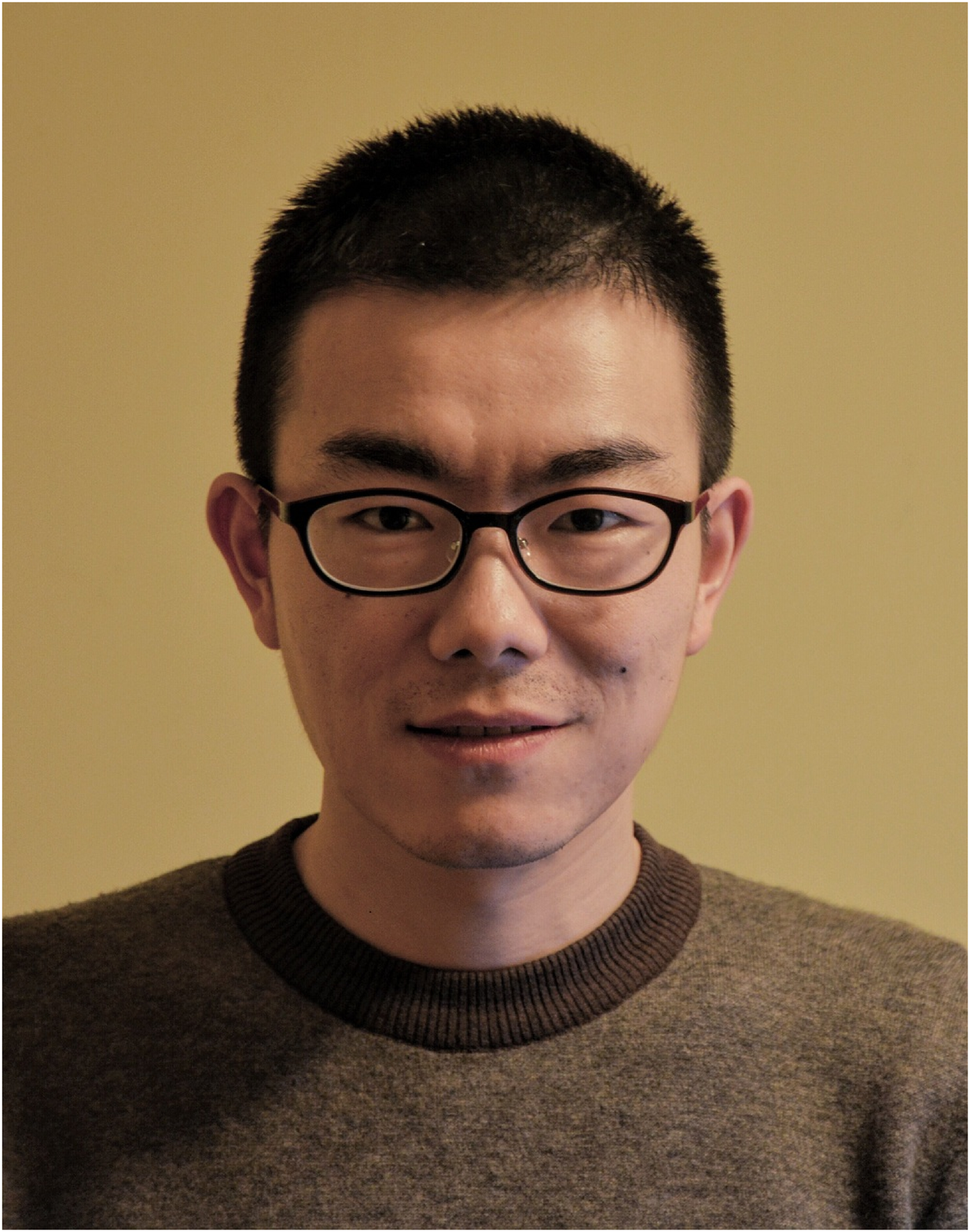}}]{Ran~Zi}
(S'14) received the B.E. degree in Communication Engineering and M.S. degree in Electronics and Communication Engineering from Huazhong University of Science and Technology (HUST), Wuhan, China in 2011 and 2013, respectively. He is currently working toward the Ph.D. degree in HUST. His research interests include MIMO systems, millimeter wave communications and multiple access technologies.
\end{IEEEbiography}

\begin{IEEEbiography}[{\includegraphics[width=1.4in,height=1.3in,clip,keepaspectratio]{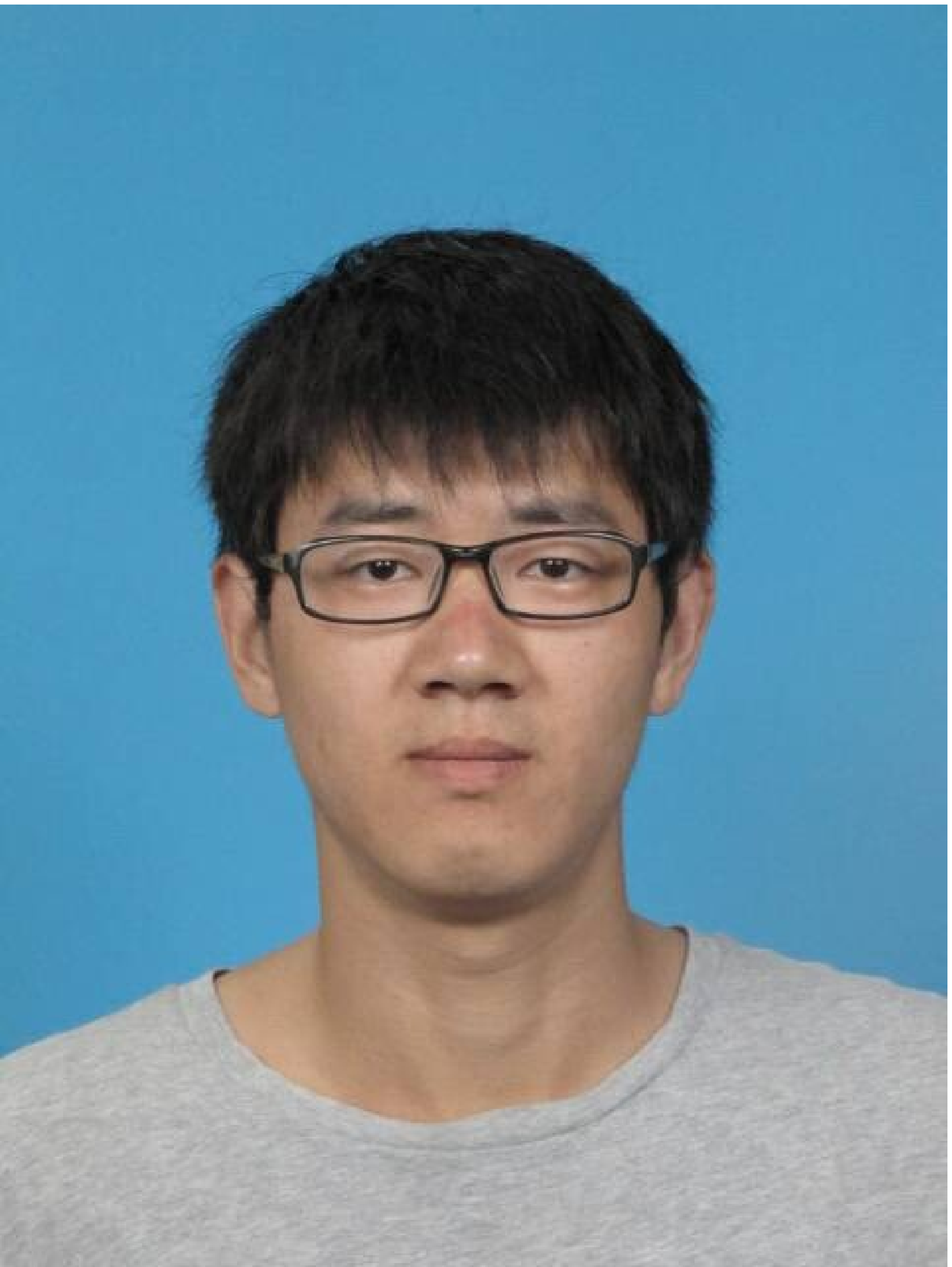}}]{Haichao Wang}
received the bachelor degree in electronic science and technology from Wuhan University of Technology, Wuhan, China, in 2013, Now he is working toward the master degree in Huazhong University of Science and Technology, Wuhan, China. His research interests include the mutual coupling effect in antenna arrays and optimization of the number of RF chains in antenna arrays.

\end{IEEEbiography}

\begin{IEEEbiography}[{\includegraphics[width=1.4in,height=1.3in,clip,keepaspectratio]{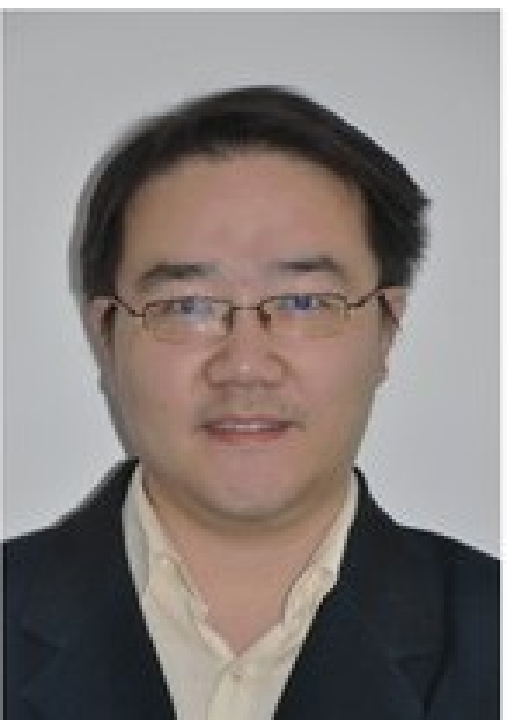}}]{Jing Zhang}
(M'13) received the M.S. and Ph.D. degrees in electronics and information engineering from Huazhong University of Science and Technology (HUST), Wuhan, China, in 2002 and 2010, respectively.
He is currently an Associate Professor with HUST. From November 2014 to November 2015, he was a Visiting Researcher with Friedrich-Alexander-University, Erlangen-Nuremberg, Germany. He has published about 20 papers in refereed journals and conference proceedings. He has done research in the areas of multipleinput multiple-output, CoMP, beamforming, and next-generation mobile communications. His current research interests include cellular systems, green communications, channel estimation, and system performance analysis.
\end{IEEEbiography}

\begin{IEEEbiography}[{\includegraphics[width=1.4in,height=1.3in,clip,keepaspectratio]{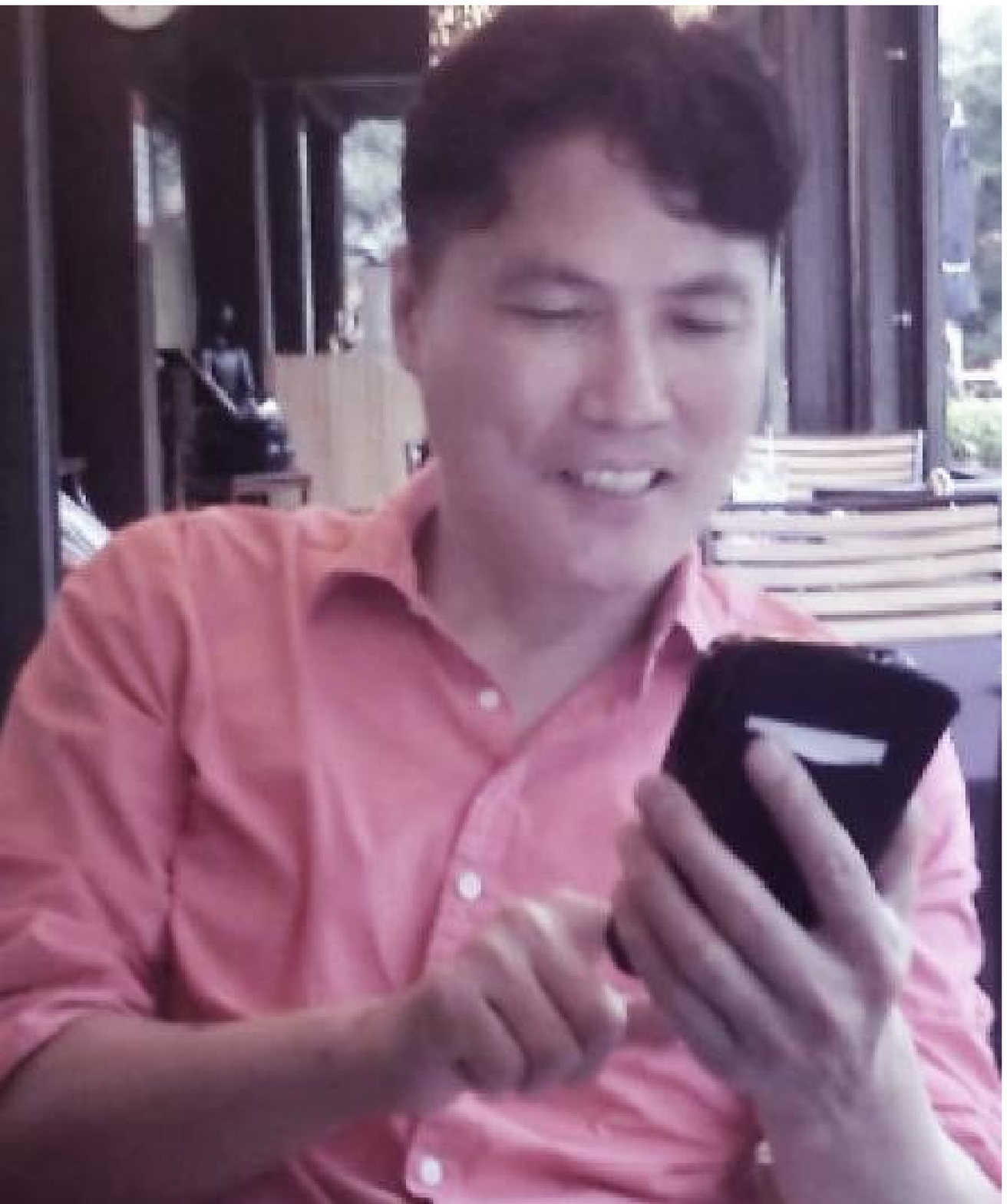}}]{Minho Jo}
(M'07) is now Professor in the Department of Computer and Information Science, Korea University, Sejong Metropolitan City, South Korea. He received his BA in the Dept. of Industrial Engineering, Chosun Univ., S. Korea in 1984, and his Ph.D. in the Dept. of Industrial and Systems Engineering, Lehigh University, USA in 1994, respectively. He is one of founders of Samsung Electronics LCD division. He was awarded with Headong Outstanding Scholar Prize with 20,000 US Dollars of prize in Dec., 2011. He is the Founder and Editor-in-Chief of the KSII Transactions on Internet and Information Systems (SCI (ISI) and SCOPUS indexed, respectively). He is currently an Editor of IEEE Wireless Communications, an Associate Editor of IEEE Internet of Things Journal, an Associate Editor of Security and Communication Networks, Associate Editor of Ad-hoc \& Sensor Wireless Networks, and an Associate Editor of Wireless Communications and Mobile Computing, respectively. He is was Vice President of the Institute of Electronics and Information Engineers (IEIE), and was Vice President of the Korea Information Processing Society (KIPS). Areas of his current interests include LTE-Unlicensed, IoT, cognitive networks, HetNets in 5G, green (energy-efficient) wireless communications, mobile cloud computing, 5G wireless communications, optimization and probability in networks, network security, software defined networks (SDN), and massive MIMO.

 \end{IEEEbiography}

\end{document}